# 2

# *Universality in the Photophysics of π-Conjugated Polymers and Single-Walled Carbon Nanotubes*

**Sumit Mazumdar, Zhendong Wang, and Hongbo Zhao**

**CONTENTS**

## 2.1 Introduction

Semiconducting carbon-based organic $\pi$-conjugated systems have been intensely investigated over the past several decades. In particular, their photophysics has been and continues to be of strong interest because of fundamental curiosity as well as current and promising technological applications. From a fundamental perspective, interest in carbon-based $\pi$-conjugated systems originates from their remarkable differences from the conventional inorganic band semiconductors. In contrast to the latter, strong, short-range, repulsive Coulomb interactions occur among the $\pi$-electrons in the organics, and these interactions contribute to a significant fraction of the optical gap. Theoretical understanding of $\pi$-conjugated systems therefore necessarily requires going beyond traditional band theory. Experimentally, even as exciton formation is found to be common in these materials, the standard technique of comparing the thresholds of linear absorption and photoconductivity for the determination of the exciton binding energy fails in noncrystalline organic materials because of the existence of disorder and inhomogeneity in these systems. Nonlinear optical spectroscopy and, in particular, ultrafast modulation spectroscopy have played valuable roles in elucidating the underlying electronic structures and photophysics of $\pi$-conjugated systems.

The goal of this chapter is to provide a theoretical background for understanding the many beautiful and sophisticated experiments being done in this area. We have chosen to describe both quasi one-dimensional $\pi$-conjugated polymers (PCPs) and semiconducting single-walled carbon nanotubes (S-SWCNTs) because of the remarkable similarities in their behavior under photoexcitation, as we describe in the following sections. The common themes between these two seemingly different classes of materials are $\pi$-conjugation, quasi one-dimensionality, and strong Coulomb interactions, and similar behavior is perhaps to be expected. Nevertheless, not all aspects of this perspective have received universal acceptance or even attention. Thus, for example, in the area of S-SWCNTs, theorists and experimentalists uniformly agree that a specific two-photon state observed above the optical exciton gives the lower threshold of the lowest continuum band. Even though ultrafast spectroscopy reveals a similar two-photon state in the PCPs, the same idea continues to meet resistance. This and other inconsistencies often lead to severe disagreements between theoretical and experimental estimations of materials' parameters such as exciton binding energies.

In the following sections we present our current understanding of the electronic structures and excited state absorptions in PCPs and S-SWCNTs within a common theoretical model. We have attempted to give complete discussions of the parameterizations of the model and the methods that have been used by various authors. We have adopted a specific configuration interaction approach for both classes of materials.

Our aim is to give physical interpretations of experiments that have been performed and to give guidance to future experimental work. One caveat is that the work reported here applies strictly to single PCP chains or single nanotubes; interchain or intertube interactions are ignored (see also Section 2.6).

## 2.2 Theoretical Model and Computational Techniques

The theoretical model that we adopt is the semiempirical $\pi$-electron approximation (Pariser and Parr 1953; Pople 1953) that is widely used to describe *planar* $\pi$-conjugated systems. The model assumes that the lowest energy excitations in planar conjugated systems involve the $\pi$-electrons only and ignores the electrons occupying orthogonal $\sigma$-bands. Because $\pi$–$\pi^*$ excitation energies decrease rapidly with system size, while the $\sigma$ and $\sigma^*$ bands are nearly dispersionless, the approximation is excellent for large systems. Thus, PCPs have been discussed extensively within the semiempirical Pariser–Parr–Pople (PPP) Hamiltonian (Pariser and Parr 1953; Pople 1953),

$$H = -\sum_{\langle ij \rangle,\sigma} t_{ij}(c^{\dagger}_{i\sigma}c_{j\sigma} + c^{\dagger}_{j\sigma}c_{i\sigma}) + U\sum_{i} n_{i\uparrow}n_{i\downarrow} + \frac{1}{2}\sum_{i\neq j} V_{ij}(n_i - 1)(n_j - 1) \tag{2.1}$$

Here,

$c^{\dagger}_{i\sigma}$ creates a $\pi$-electron with spin $\sigma(\uparrow, \downarrow)$ in the $p_z$ orbital of the $i$th carbon atom;
$\langle ij \rangle$ implies nearest neighbor (n.n.) atoms $i$ and $j$;
$n_i\sigma = c^{\dagger}_{i\sigma}c_{j\sigma}$ is the number of $\pi$-electrons with spin $\sigma$ on the atom $i$;
$n_i = \Sigma\sigma\, n_i\sigma$ is the total number of $\pi$-electrons on the atom;
the parameter $t_{ij}$ is the one-electron hopping integral between $p_z$ orbitals of n.n. carbon atoms;
$U$ is the on-site electron–electron (e–e) repulsion between two $\pi$-electrons occupying the same carbon atom $p_z$ orbital; and
$V_{ij}$ is the intersite e–e interaction.

Limiting the electron hopping to nearest neighbors does not lead to loss of generality. In Section 2.5.2.4 we discuss the role of more distant hopping.

In recent years, considerable work has also been done on PCPs as well as S-SWCNTs within ab initio approaches. The procedure here consists of obtaining the ab initio ground state, which is followed by the determination of the quasi-particle energies within the GW approximation and the solution of the Bethe–Salpeter equation of the two-particle Green's function (Rohlfing and Louie 1999; van der Horst et al. 2001; Puschnig and Ambrosch-Draxl 2002; Ruini et al. 2002; Spataru et al. 2004; Chang et al. 2004; Perebeinos, Tersoff, and Avouris 2004). This approach avoids the obvious disadvantages associated with assuming $\sigma$–$\pi$ separation (this can be of serious concern in the context of S-SWCNTs, especially for the narrow ones) as well as choosing seemingly arbitrary parameters.

However, serious disagreements exist between the predictions of the ab initio and the semiempirical theories (Wang, Zhao, and Mazumdar 2006), and experimental results of ultrafast spectroscopy and other nonlinear optical measurements overwhelmingly agree with the predictions of the latter (Zhao et al. 2006). We speculate that this might be due to the notorious difficulty associated with taking the on-site Coulomb repulsion $U$ into account within ab initio approaches. On the other hand, many-body problems that would be formidable within the ab initio approach—such as the enhancement of the ground state bond alternation in polyacetylene by e–e interactions (Baeriswyl, Campbell, and Mazumdar 1992) or the occurrence of the lowest two-photon state below the optical state in the same system (Soos, Ramasesha, and Galvão 1993)—can be understood relatively easily within Equation (2.1). We will therefore limit our discussions to within the semiempirical model.

Equation (2.1) does not include electron–phonon interactions, which are known to have strong effects in low-dimensional systems. This is primarily because we will be interested in high-energy excited states, which are difficult to investigate within models that incorporate both electron–electron and electron–phonon interactions. Therefore, our results pertain primarily to the rigid bond approximation. We cite theoretical and experimental results, however, that are distinct consequences of electron–phonon interactions, wherever appropriate.

Even within the rigid bond $\pi$-electron approximation, the many-body nature of Equation (2.1) makes computations of electronic structures and linear and nonlinear absorptions extremely difficult. Three distinct approaches have been popular over the years. We briefly discuss each of these next.

*Exact diagonalization.* This approach is identical to full configuration interaction (FCI) and is obviously the most accurate. Unfortunately, the total number of configurations increases roughly as $4^N$, where $N$ is the number of carbon atoms, and computation of excited state properties is currently limited to $N = 12$ (McWilliams, Hayden, and Soos 1991). (Ground state properties with special symmetries have been investigated up to $N = 16$; see Li, Mazumdar, and Clay.) Clearly, this approach is suitable only for linear polyenes and finite oligomers of polydiacetylene (PDA) (McWilliams et al. 1991). When accompanied with finite size scaling ($N \to \infty$ extrapolations), the method can give very useful information on the energetics of the lowest excited states of *trans*-polyacetylene ($t$-$(CH)_x$) and the PDAs (Ramasesha and Soos 1984). In general, however, even for $t$-$(CH)_x$ and PDA, it is difficult to obtain information on the very high energy excited states that are relevant in ultrafast spectroscopy within the standard exact diagonalization schemes based on the usual configuration-space or molecular orbital basis functions. We will report the results of a different approach to exact diagonalization here that focuses on the *wavefunctions,* rather than energetics. Pictorial and physical characterizations of excited states that are most relevant to the photophysics are obtained within this exciton basis valence bond description. It is believed that these physical characterizations apply to the long chain limit.

*Finite-order configuration interaction.* This rather broad class of techniques includes the single-CI (SCI), higher order CI such as single- and double-CI (SDCI) and quadruple-CI (QCI), the multiple-reference single- and double-CI (MRSDCI) and the coupled-cluster approach. The SCI, which includes the CI between only singly excited configurations from the Hartree–Fock (HF) ground state, has been widely applied to PCPs (Abe et al. 1992; Yaron and Silbey 1992; Gallagher and Spano 1994; Chandross et al. 1997; Chandross and Mazumdar 1997) and, more recently, to S-SWCNTs (Zhao and Mazumdar 2004; Zhao et al. 2006; Wang et al. 2006; Wang, Zhao, and Mazumdar 2007). The problems with this approach are well known. Two-photon spin singlet excitations that are superpositions of two or more local triplets (such as the $2^1A_g$ in linear polyenes) cannot be described within the SCI (Hudson, Kohler, and Schulten 1982; Ramasesha and Soos 1984). On the other hand, higher order CIs are size inconsistent (with the possible exception of the MRSDCI) and also quickly become impossible to implement as the system size increases. We will show that with proper parameterization, the SCI can give reasonably accurate descriptions of eigenstates that are predominantly one-electron–one-hole (1e–1h) relative to the ground state and that lie within a specific energy range (Chandross and Mazumdar 1997). The MRSDCI (Tavan and Schulten 1987; Shukla, Ghosh, and Mazumdar 2003) and the coupled-cluster (Shuai et al. 2000) approaches have also been used to understand specific excited states.

*Density matrix renormalization group.* The DMRG is a highly accurate many-body method that is particularly suitable for one-dimensional systems (White 1992). Two different groups have applied this technique extensively to $t$-$(CH)_x$, PDAs, and polyphenylenes (Shuai et al. 1997; Pati et al. 1999; Lavrentiev et al. 1999; Ramasesha et al. 2000; Bursill and Barford 2002; Race, Barford, and Bursill 2003). A specific, narrow S-SWCNT has also been investigated within this approach (Ye et al. 2005). Although the method targets specific excited states and can, in principle, obtain their energies in the long chain limit, it appears that partial information about these excited states should already be available through other means. Whether or not the long-range part of the Coulomb interaction in Equation (2.1) is incorporated accurately is also controversial (Ramasesha et al., 2000, for example, have performed calculations only with n. n. intersite Coulomb interactions). In any event, the DMRG is unsuitable for S-SWCNTs in the diameter region of interest (~1 nm), so we will discuss it no further.

Among the preceding approaches, only the SCI is suitable for S-SWCNTs. Because the principal goal of our work is to demonstrate the universal photophysics of PCPs and S-SWCNTs, we will focus on this technique. However, the SCI misses important excited states, so it needs to be justified and should be applied with caution. With this in mind, we proceed to discuss the photophysics of polyacetylenes and PDAs in the next section.

## 2.3 Polyacetylenes and Polydiacetylenes

The exciton behavior of PDAs with crystalline forms has been investigated already in the early 1980s by many groups (Chance et al. 1980; Sebastian and Weiser 1981; Tokura et al. 1984; Kajzar and Messier 1985). Nevertheless, it is probably fair to say that modern exciton physics of PCPs began with the observation by Fann et al. (1989) that a two-photon $^1A_g$ state nearly degenerate with the optical $1^1B_u$ state in *t*-$(CH)_x$ exists. This conclusion was arrived at from the energy locations of three- and two-photon resonances in third harmonic generation (THG) measurement in *t*-$(CH)_x$ (Fann et al. 1989). THG measurement in a derivative of *t*-$(CH)_x$, poly(1,6)-heptadiyne, led to the same conclusion (Halvorson et al. 1993). These observations were taken to be direct proofs for weak Coulomb interactions in PCPs other than PDAs because the $2^1A_g$ is degenerate with the $1^1B_u$ in the long chain limit of linear polyenes in the $U = 0$ limit of Equation (2.1).

The broader experimental scenario was highly confusing, however, for two reasons. First, it had been known for more than a decade that the $2^1A_g$ in finite polyenes occurred below the $1^1B_u$ and that the separation between the $1^1B_u$ and the $2^1A_g$ increased with increasing chain length (Hudson et al. 1982). Second, numerical simulations of the THG spectra within one-electron Hückel theory by different groups failed to reproduce the two-photon resonance due to the $2^1A_g$, even as a very prominent three-photon resonance due to the $^1B_u$ was obtained (Yu et al. 1989; Wu and Sun 1990; Shuai and Brédas 1991; Halvorson et al. 1993). The absence of the two-photon resonance to the $2^1A_g$ was shown to be due to cancellation between nonlinear optical channels making positive and negative contributions to third-order optical susceptibility (Guo, Guo, and Mazumdar 1994; Mazumdar and Guo 1994). Importantly, this last work also showed that such cancellations were incomplete, and therefore two-photon resonances were visible only when the excitation spectrum was excitonic. This theoretical result clearly indicates the inapplicability of one-electron theories to PCPs.

Two parallel developments gave the hint toward the correct interpretation of the THG resonances in *t*-$(CH)_x$. It was observed from SDCI (Heflin et al. 1988) and exact diagonalization (Soos and Ramasesha 1989) calculations for finite polyenes that, even within the PPP Hamiltonian (Equation 2.1), an excited $^1A_g$ state exists *above* the $1^1B_u$ with unusually large transition dipole coupling with the $1^1B_u$. Dipole couplings between all other excited $^1A_g$ states and the $1^1B_u$ were one to two orders of magnitude smaller. Based on very similar calculations, Dixit, Guo, and Mazumdar (1991) claimed that this result was independent of Coulomb parameters and that the specific $^1A_g$ state, which they labeled the $m^1A_g$, was energetically close to the $1^1B_u$ in the long chain limit of linear polyenes and was the origin of the two-photon resonance in THG. In subsequent work, this particular research group claimed that the $1^1B_u$ and the $m^1A_g$ were excitons and, together with the $n^1B_u$, the threshold state of the continuum band, formed the "essential states" that determined third-order optical nonlinearity of polyacetylenes and PDAs (Guo et al. 1993).

The preceding calculations were exact but for finite polyenes. SCI calculations by several groups, performed about the same time found related results for the long chain limit: (1) the $1^1B_u$ is an exciton, and (2) a specific $^1A_g$ state exists above the $1^1B_u$ but below the conduction band threshold that dominates third-order optical nonlinearity (Abe et al. 1992; Yaron and Silbey 1992; Gallagher and Spano 1994). This apparent similarity in the exact results (Dixit et al. 1991; Guo et al. 1993) and SCI (Abe et al. 1992; Yaron and Silbey 1992; Gallagher and Spano 1994) was comforting, but still required further work that would explain the origin of this similarity and provide a physical interpretation of the results (in particular because SCI and exact calculations in all cases differed on the location of the $2^1A_g$ state relative to the $1^1B_u$). With this motivation, the exciton-basis valence bond (VB) approach (Chandross, Shimoi, and Mazumdar 1999) was developed. A detailed discussion of this approach and the photophysics of polyacetylenes and PDAs within this theory follows; this justifies careful application of the SCI for understanding the photophysics of PCPs.

### 2.3.1 The Exciton-Basis VB Theory

The exciton-basis VB theory is an FCI approach that focuses on wavefunctions, rather than energetics. It recognizes at the outset that understanding of the photophysics of PCPs within the VB exact diagonalization (Soos and Ramasesha 1989; Dixit et al. 1991) or CI approaches using a molecular orbital (MO) basis is difficult because, in both cases, for intermediate *U* the wavefunctions are superpositions of so many configurations that simple physical interpretation is lost. Although the MO theory is valid for the small-*U* regime, VB theory is valid for very large *U*. The exciton VB basis is a hybrid of MO and VB bases and is particularly suitable for obtaining physical interpretations of excitations at intermediate *U*.

Within the VB exciton basis representation, we consider a polyacetylene chain as coupled ethylenic units. The PPP Hamiltonian can then be rewritten as (Mazumdar and Chandross 1997; Chandross et al. 1999)

$$H = H_{\text{intra}} + H_{\text{inter}}$$
$$H_{\text{intra}} = H_{\text{intra}}^{\text{ee}} + H_{\text{intra}}^{\text{CT}} \quad (2.2)$$
$$H_{\text{inter}} = H_{\text{inter}}^{\text{ee}} + H_{\text{inter}}^{\text{CT}}$$

where $H_{\text{intra}}$ and $H_{\text{inter}}$ describe the interactions within one ethylenic unit and the interactions between units, respectively. Each of these two terms contains two parts: the e–e interactions and the charge transfer (CT) or the electron hopping. $H_{\text{intra}}^{\text{CT}}$ describes the electron hopping within one unit. Its eigenstates simply correspond to the bonding and antibonding MOs of the unit:

$$a_{i,\lambda,\sigma}^{\dagger} = \frac{1}{\sqrt{2}}\left[c_{2i-1,\sigma}^{\dagger} + (-1)^{\lambda-1} c_{2i,\sigma}^{\dagger}\right] \quad (2.3)$$

where $a_{i,\lambda,\sigma}^{\dagger}$ creates an electron of spin $\sigma$ in the bonding ($\lambda = 1$) or the antibonding ($\lambda = 2$) MO of ethylene unit *i*. $H_{\text{intra}}^{\text{ee}}$ introduces CI between configurations within a unit. $H_{\text{inter}}^{\text{CT}}$ contains three terms (Mazumdar and Chandross 1997; Chandross et al. 1997) corresponding to the electron hopping among (1) the bonding MOs of the neighboring units, (2) the antibonding MOs of the neighboring units, and (3) the bonding MO of one unit and the antibonding MO of a neighboring unit. These electron hoppings of $H_{\text{inter}}^{\text{CT}}$ are illustrated in Figure 2.1(a) by arrow-headed lines. $H_{\text{intra}}^{\text{ee}}$ also contains three different terms (Mazumdar and Chandross 1997; Chandross et al. 1997) corresponding to (1) density-density correlations, such as static Coulomb interactions between electrons within the same or different MOs; (2) products of density and electron hopping between MOs; and (3) products of two hopping terms.

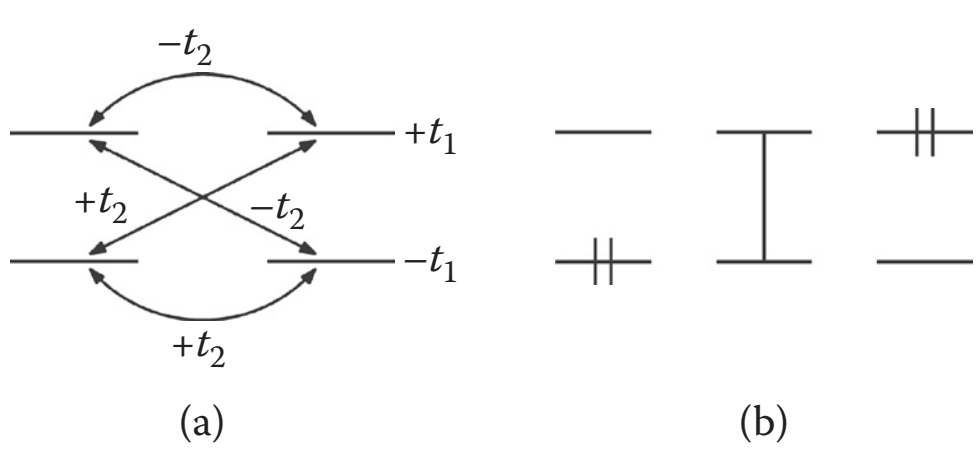

**FIGURE 2.1**
(a) Electron transfers induced by $H_{\text{intra}}^{\text{CT}}$. $t_1$ and $t_2$ are the intra- and interunit hopping integrals, respectively. (b) The VB exciton basis diagrams for one ethylene unit. Bonding and antibonding MOs of the two-level system are occupied by zero, one, and two electrons.

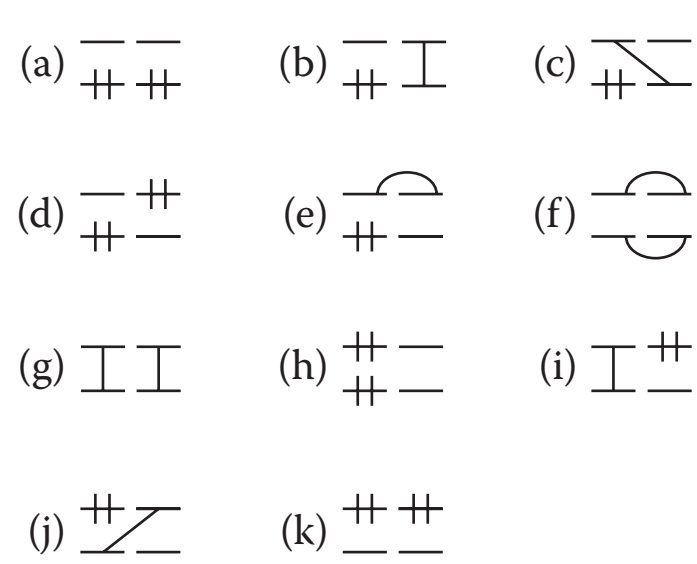

**FIGURE 2.2**
Exciton-basis diagrams for a two-unit oligomer. Singly occupied MOs are paired as singlet bonds. Mirror-plane and charge-conjugation symmetries are assumed. (From Chandross, M. et al., *Phys. Rev. B*, 59, 4822, 1999. With permission.)

The VB exciton basis is best understood pictorially. The following convention has been adopted for the exciton basis diagrams. A line denoting a singlet bond is constructed between spin-bonded singly occupied MOs because there is equal probability that each MO is singly occupied by an up or down spin. The VB exciton basis for ethylene is trivial and consists of only three diagrams: (1) doubly occupied bonding MO and empty antibonding MO, (2) singly occupied bonding and antibonding MOs, (3) empty bonding MO and doubly occupied antibonding MO, as illustrated in Figure 2.1(b).

We go beyond the one-unit case and illustrate the exciton basis for the two-unit case in Figure 2.2 (Mazumdar and Chandross 1997; Chandross et al. 1999). $C_{2h}$ as well as charge-conjugation symmetry (CCS) are assumed in this figure; therefore, a single diagram is used to represent the full set of diagrams related by $C_{2h}$ and/or CCS. The diagram (a) in Figure 2.2 is the product wavefunction of the ground states of two noninteracting units. The nature of the corresponding "ground state" diagram is the same for the $N$-unit chain with $N >> 2$ (Mazumdar and Chandross 1997; Chandross et al. 1999). All many-electron diagrams for the $N$-unit chain with two or fewer excitations can be constructed by taking the direct product of the ($N-2$)-unit ground state diagram and the one- or two-electron excitations of Figure 2.2, with the understanding that the locations of the electrons and holes are now arbitrary.

Chandross and Mazumdar (1997) and Chandross et al. (1999) have shown that, within the VB exciton basis, the nonlinear optical channels can be summarized schematically as in Figure 2.3, which is valid for intermediate coupling regime appropriate for PCPs. Because the exact solution for larger systems turned out to be very difficult, they chose an $N = 5$ finite polyene chain. From the wavefunctions of the exact solution, the schematic Figure 2.3 was constructed.

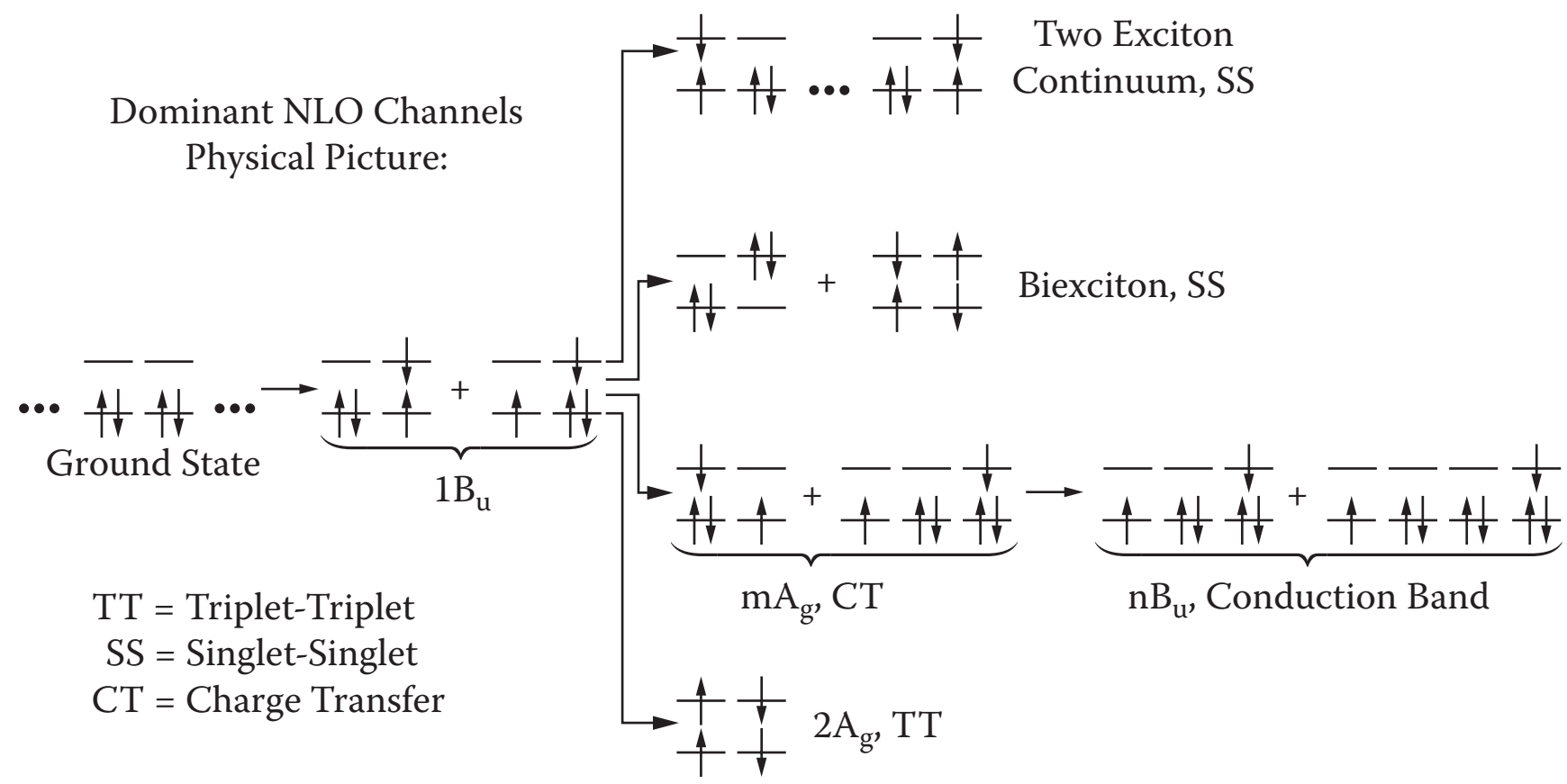

**FIGURE 2.3**
The dominant nonlinear optical channels in *t*-$(CH)_x$. (From Chandross, M. et al., *Phys. Rev. B,* 59, 4822, 1999. With permission.)

The nature of the transition dipole coupling operator only allows the optical excitations between states with different parities, $A_g \leftrightarrow B_u$. In Figure 2.3, optical absorption from the ground state $1^1A_g$ leads to the $1^1B_u$ exciton, which is dominated by single-electron excitations, both within a unit as well as with short-distance CT. There are multiple options for the second step in the optical process. The first possibility is returning to the ground state. Other than that, the creation of a second excitation on the same or a neighboring unit can lead to the two-triplet $2^1A_g$ or to the biexciton state, while a second excitation far from the first gives the threshold of the two-exciton continuum. In addition to the preceding processes, further CT from the $1^1B_u$ single-electron excitation can occur, leading to a state with greater electron–hole separation than in the $1^1B_u$. Taken together with calculations of transition dipole coupling, it was shown that this single-excitation state with greater electron–hole separation was the $m^1A_g$. It was also shown that the $m^1A_g$ was an exciton. Further absorption and charge separation from the $m^1A_g$ leads to the $n^1B_u$ state, which was shown to constitute the threshold state of the continuum band from calculations of electroabsorption.

### 2.3.2 Justification of the SCI

The unit cells of most PCPs are much larger than those of *t*-$(CH)_x$, and FCI is consequently impossible for these systems. Figure 2.3 gives a strong justification for the use of the SCI in these cases. As seen in the figure, the dominant nonlinear optical channel consists of excitations ($1^1B_u$, $m^1A_g$, and $n^1B_u$), all of which are *single excitations* with respect to the correlated ground state. Thus, even though the SCI will miss the triplet–triplet and two-exciton excitations, with *proper parameterization of the Hamiltonian*, it should be possible to describe accurately the limited energy space between the optical exciton and the lower threshold of the continuum band. We show in the next section that this is indeed true.

## 2.4 Poly-paraphenylenes and Poly-paraphenylenevinylenes

### 2.4.1 SCI and Parameterization of the PPP Hamiltonian

Poly-paraphenylene (PPP) and poly-paraphenylenevinylene (PPV), together with their derivatives, have been investigated almost as intensely as $t$-$(CH)_x$ and PDAs. It was recognized simultaneously by several groups (Rice and Gartstein 1994; Cornil et al. 1994; Chandross et al. 1994) that the absorption spectra of these systems with multiple absorption bands allow unambiguous demonstration of the strong role of e–e interactions (as opposed to the polyacetylenes and PDAs with a single optical absorption, which can always be fit by simply varying the multiple electron–electron or electron–phonon interaction parameters that exist in the literature). Because very similar arguments have recently been used also in the context of S-SWCNTs (see Section 2.5.2), we reproduce them here.

The unit cell of PPV has eight carbon atoms, which, within one-electron theory, gives four valence bands and four conduction bands. Of the four valence bands, three are extended or delocalized over the entire polymer, and one is localized, with zero electron densities on the para-carbons of the phenyl unit and the vinyl carbons. We label the delocalized valence bands as $d_1$, $d_2$, and $d_3$, respectively (with $d_1$ the closest to the chemical potential and $d_3$ the farthest) and the localized valence band as $l$ (Chandross et al. 1994). The $l$-band is the second band from the chemical potential. We label the corresponding conduction bands as $d_1^*$, $d_2^*$, $d_3^*$, and $l^*$. In the case of PPP, the outermost bands $d_3$ and $d_3^*$ are missing, but the band structure otherwise is similar. The $d_2 \rightarrow d_2^*$ and $d_3 \rightarrow d_3^*$ energy gaps are very large, even within one-electron theory, and are outside the experimental range. The relevant optical excitations are then the symmetric $d_1 \rightarrow d_1^*$ and $l \rightarrow l^*$ excitations and the asymmetric, degenerate excitations $d_1 \rightarrow l^*$ and $l \rightarrow d_1^*$.

From explicit calculations of transition dipole moments, the symmetric and asymmetric excitations are polarized predominantly (exactly) parallel and perpendicular to the polymer axes in PPV (PPP). The degenerate $d_1 \rightarrow l^*$ and $l \rightarrow d_1^*$ excitations occur at an energy exactly in the middle of the $d_1 \rightarrow d_1^*$ and $l \rightarrow l^*$ energy gaps within one-electron theory, which then predicts a highly symmetric absorption band spectrum, with low- and high-energy absorption bands polarized along the longitudinal direction and a weaker transverse band exactly in the center of these two bands. Repeated experiments by several groups (Chandross et al. 1997; Comoretto et al. 1998, 2000; Miller et al. 1999) have shown that although the longitudinal absorption bands in PPV and its derivatives occur at 2.2–2.4 eV and at 6.0 eV, the transverse absorption occurs at ~4.7 eV. The transverse absorption is therefore blue shifted from its expected location by ≥0.5 eV.

The qualitative explanation for this is simple within the many-electron PPP Hamiltonian. Specifically, the matrix element $\langle \psi_{d_1\to l^*} | H_{ee} | \psi_{l\to d_1^*} \rangle$ is nonzero, and as a consequence eigenstates of the Hamiltonian with nonzero e–e interactions are odd and even superpositions of these basis functions, $\Psi_O = \psi_{d_1\to l^*} - \psi_{l\to d_1^*}$ and $\Psi_E = \psi_{d_1\to l^*} + \psi_{l\to d_1^*}$. Repulsive $H_{ee}$ ensures that the optically forbidden $\Psi_O$ is red shifted while the optically allowed $\Psi_E$ is blue shifted. The observed 4.7-eV absorption band within this picture is simply the allowed transition from the ground state to $\Psi_E$.

Note that the preceding qualitative discussion is entirely within the SCI. In the following we make this quantitative and arrive at the correct parameterization of the PPP parameters by insisting that the Hamiltonian should yield a calculated absorption spectrum that matches the experimental spectrum. Our discussion is based on the procedure used by Chandross and Mazumdar (1997) and Chandross et al. (1997).

The hopping integrals chosen by Chandross and Mazumdar were the standard ones: $t_{ij} = t = 2.4$ eV between the phenyl carbon atoms and $t_{ij} = t_1$ ($t_2$) = 2.2 (2.6) eV for the single (double) bonds of the vinylene linkage. The particular form for the long-range Coulomb interactions chosen by these authors was

$$V_{ij} = \frac{U}{\kappa\sqrt{1 + 0.6117 R_{ij}^2}} \tag{2.4}$$

$R_{ij}$ is the distance between atoms $i$ and $j$ in angstroms and $\kappa$ is an effective dielectric constant. Within the standard Ohno (1964) parameterization of the Coulomb potential, $U$ = 11.13 eV and $\kappa$ = 1. Unlike the Ohno parameterization, however, Chandross and Mazumdar (1997) considered five values of $U$ = 2.4, 4.8, 6, 8, and 10 eV and three values of $\kappa$ = 1, 2, and 3. Only with $U$ = 8 eV and $\kappa$ = 2 could they reproduce the experimental absorption spectrum of MEH-PPV. The experimental absorption spectrum here and the theoretical fit calculated for an eight-unit PPV oligomer with these $U$ and $\kappa$ are shown in Figure 2.4(a). For all other $U$ and $\kappa$, the calculated absorption spectra were remarkably different (for further details see Table I in Chandross et al. 1997).

### 2.4.2 Ultrafast Spectroscopy of PPV Derivatives

We have used the same interaction parameters to obtain quantitative descriptions of nonlinear absorption in PPV derivatives with weak

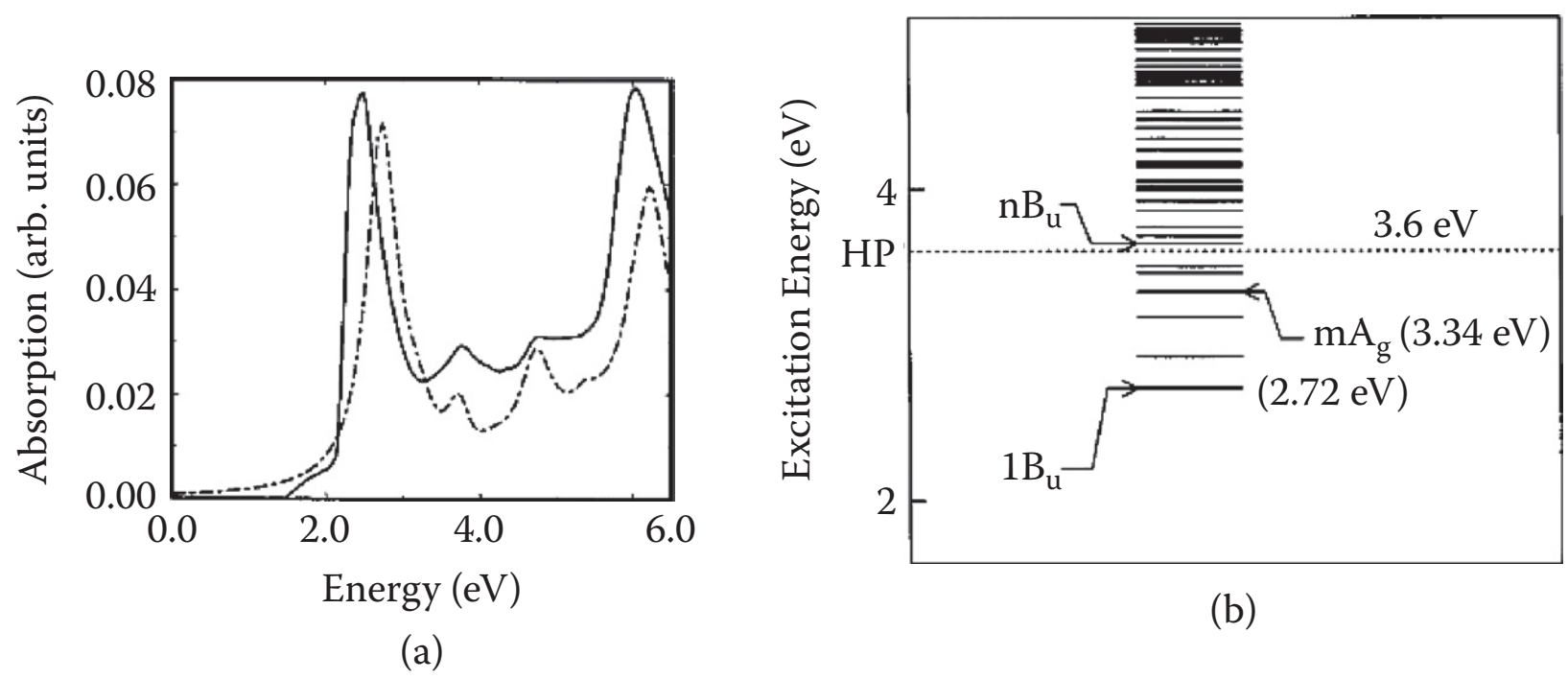

**FIGURE 2.4**
(a) Experimental absorption spectrum of (poly[2-methoxy-5-(2′-ethyl)hexyloxy-1,4-phenylene vinylene) (MEH-PPV) (solid line) and the calculated absorption spectrum (dashed line) for an eight-unit oligomer with $U$ = 8 eV and $\kappa$ = 2. (b) The energy spectrum of an eight-unit oligomer of PPV with $U$ = 8 eV and $\kappa$ = 2. (From Chandross, M. and Mazumdar, S., *Phys. Rev. B,* 55, 1497, 1997. With permission.)

interchain interactions. In Figure 2.4(b) we show the calculated energy spectrum of an eight-unit PPV chain with $U$ = 8 eV and $\kappa$ = 2. The $1^1B_u$ is the lowest optical exciton. In SCI theory, the continuum threshold is at the HF band gap, which is also included in the figure. The continuum band threshold was obtained also with a second approach that utilizes the information on excited eigenstates in Figure 2.4(b)—namely, that the $m^1A_g$ has unusually large dipole coupling with both the $1^1B_u$ and the $n^1B_u$. Calculations of transition dipole moments then allow us to determine the $n^1B_u$ state in a two-step procedure: We determine the $m^1A_g$ from calculations of transition dipole moments with the $1^1B_u$ and then determine the $n^1B_u$ from calculations of transition dipole moments with the $m^1A_g$. We have shown the calculated $n^1B_u$ state also in Figure 2.4(b), where it is seen to be very close to the HF band gap. The calculated energies of the $n^1B_u$ (3.65 eV) and the $1^1B_u$ (2.7 eV) are both slightly too high. Our interest lies, however, in the difference between these two energy states and, even with the uncertainties associated with the SCI approximation, we see that the predicted exciton binding energy is ~0.9 ± 0.2 eV.

The $m^1A_g$ is predicted to be then much further from the $1^1B_u$ in PPV than in $t$-$(CH)_x$. This state should be visible in THG, two-photon absorption (TPA) and ultrafast photoinduced absorption (PA). In Figure 2.5(a) we show the transient photomodulation (PM) spectrum of dioctyloxy-PPV (DOO-PPV) (Frolov et al. 2000). Two different PA bands are seen in PA: a low-energy $PA_1$ band with a threshold at ~0.7 eV and a peak at ~0.9 eV and high-energy weaker $PA_2$ at ~1.4 eV. $PA_1$ is ascribed to excited state absorption to the $m^1A_g$ from the $1^1B_u$. This is in good agreement with the theoretical prediction of a gap of 0.7 eV between the $m^1A_g$ and the $1^1B_u$ (see Figure 2.4b). Strong TPA to the $m^1A_g$, as well as the detailed dynamics of $PA_1$ (Frolov et al. 2000), have confirmed this assignment. Figure 2.5(b) shows the transient PM spectrum for poly(9,9-dioctylfluorene) (PFO), another PCP that has similar basic structure to that of PPP. The theory of PPV also applies to PFO, as shown in the similar PM spectra. We will discuss Figure 2.5(c) in Section 2.5.4.

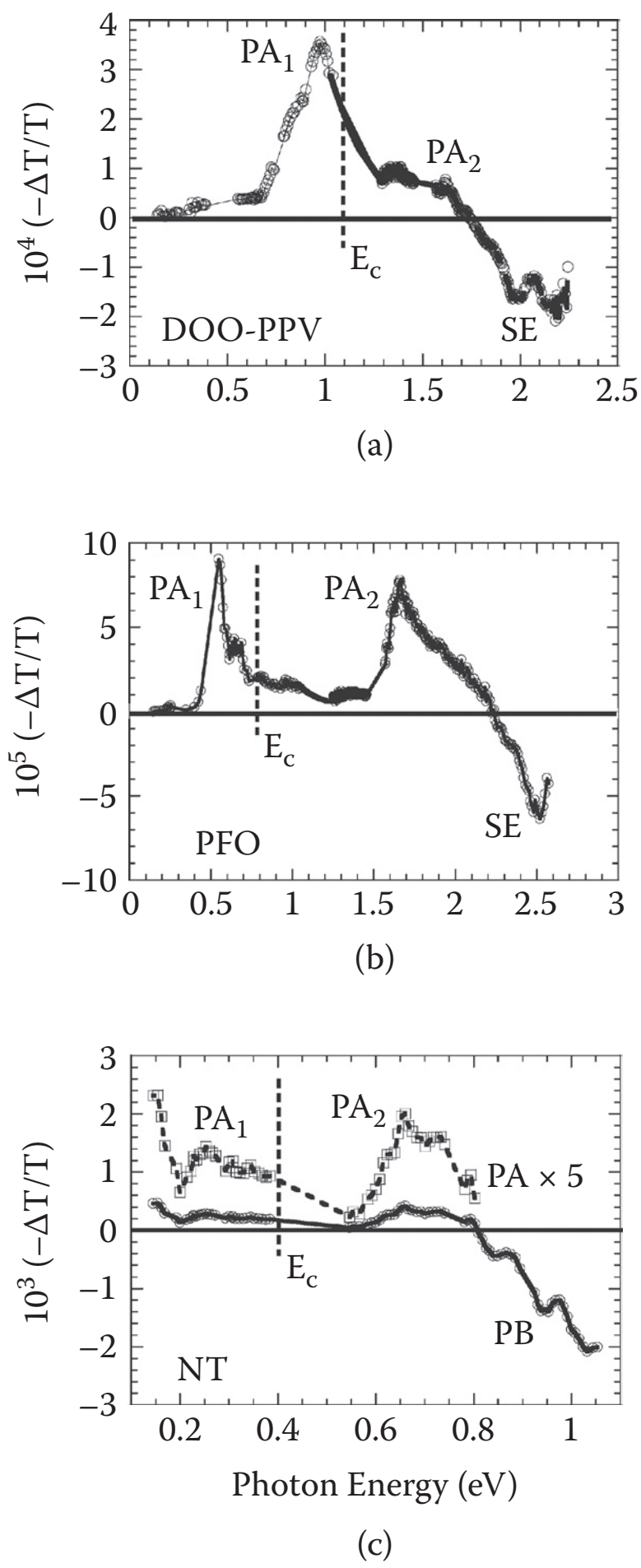

**FIGURE 2.5**
Transient PM spectra at $t = 0$ of films of (a) DOO-PPV, (b) PFO, and (c) isolated SWCNT in PVA matrix. Various PA, PB, and SE bands are assigned. The vertical dashed lines at $E_c$ between $PA_1$ and $PA_2$ denote the estimated continuum band onset. (From Zhao, H. et al., *Phys. Rev. B*, 73, 075403, 2006. With permission.)

$PA_2$ has been ascribed to a yet higher $^1A_g$ state that has been labeled the $k^1A_g$ (Frolov et al. 2000). Importantly, the $k^1A_g$ has also been seen in TPA, confirming this characterization (Frolov et al. 2000). The relaxation dynamics of the $k^1A_g$ is, however, very different from that of the $m^1A_g$. Instead of relaxing back to the $1^1B_u$, the $k^1A_g$ undergoes charge separation, suggesting that its nature is different. This has been confirmed from very detailed MRSDCI calculations of long PPV and PPP oligomers by Shukla et al. (2003). We do not discuss this further because this will take us substantially away from the principal thesis of the present work. Instead, we show in Figure 2.6 our characterizations of the $m^1A_g$ and the $k^1A_g$ within the MRSDCI. As seen in Figure 2.6, the $k^1A_g$ is expected only in systems with multiple valence and conduction bands. This explains naturally the absence of this feature in *t*-$(CH)_x$ and PDAs.

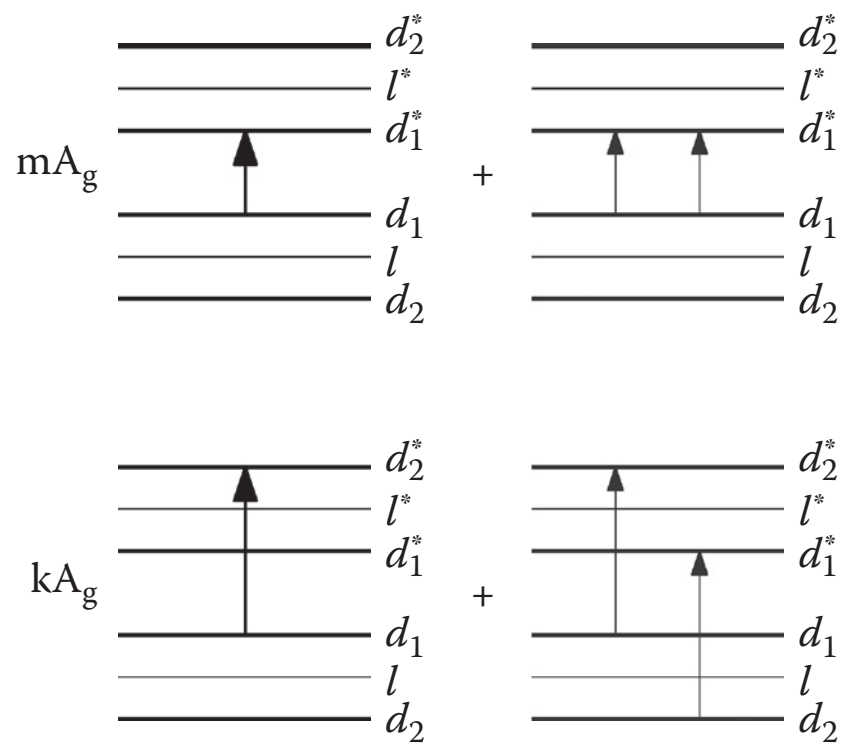

**FIGURE 2.6**
Schematic representations of the $m^1A_g$ and $k^1A_g$ states for PPP and PPV. Thicker horizontal lines imply bands with finite widths. The $m^1A_g$ is a superposition of singly and doubly excited configurations, both of which involve predominantly $d_1$ and $d_1^*$ bands only. The one excitation component of $k^1A_g$ involves $d_1 \rightarrow d_2^*$ excitations in PPP ($d_1 \rightarrow d_3^*$ in PPV), while the two-excitation components involve the $l$ and $l^*$ bands. The thick arrows denote single excitations with $^1A_g$ symmetry reached by two successive dipole-allowed transitions. Wherever applicable, particle–hole reversed excitations are also implied.

## 2.5 Semiconducting Single-Walled Carbon Nanotubes

The similarity in the photophysics of PCPs and S-SWCNTs was recognized and established through a series of papers by Mazumdar and co-workers (Zhao and Mazumdar 2004, 2007; Zhao et al. 2006; Wang et al. 2006; Wang, Zhao, et al. 2007) and, more recently, by Tretiak (2007) and Scholes et al. (2007). The strongest effect of e–e interactions is on eigenstates that are degenerate in the one-electron limit. Such degeneracies are lifted by Coulomb interaction. The strong blue shift of the allowed optical absorption transverse to the polymer axis in PPV, discussed in Section 2.4.1, is a key example of such lifting of degeneracy and provides a measure of the strength of the Coulomb interactions among the $\pi$-electrons. Our experience with PPV allowed us to make related predictions in S-SWCNTs in the context of transverse optical excitons as well as longitudinal dark excitons (Zhao and Mazumdar 2004).

Because of carbon nanotubes' nonplanar nature, the hopping integral in S-SWCNTs has a value different from those in the PCPs. We begin our discussions with the parameterization of the hopping integral in Section 2.5.1. In Section 2.5.2 we present our theoretical results for optical absorptions polarized transverse to the nanotube axes and compare these with available experimental data. Following this, in Section 2.5.3 we discuss longitudinal excitons, the splitting of bright and dark excitons, their binding energies, and the overall excitonic spectra. We show that calculations within the $\pi$-electron Hamiltonian with only two free parameters can quantitatively reproduce known experimental quantities for a large number of S-SWCNTs with diameters within a certain range and have often led to theoretical predictions that were confirmed experimentally only later. In Section 2.5.4 we briefly discuss ongoing and future work in the broad area of S-SWCNT photophysics and related topics.

### 2.5.1 Parameterization and Boundary Conditions

Because the interaction among the $\pi$-electrons depends only on the distance between them and not on the spatial topology of carbon atoms, it is logical to assume that $U$ and $V_{ij}$ are exactly the same as in the PCPs. We use Equation (2.4) with $U = 8$ eV and $\kappa = 2$ in all our calculations of S-SWCNTs.

For the n.n. electron hopping parameter $t$, topology does make a difference. Unlike most PCPs that are planar, carbon atoms of SWCNTs are on a cylinder, and this curvature implies smaller $p_z$–$p_z$ orbital overlap between n.n. carbon atoms and hence a smaller $t$. In the same spirit of fitting $U$ and $V_{ij}$ for PPV (see Section 2.4.1), Wang and colleagues (2006) arrived at the proper $t$ for S-SWCNTs by fitting the calculated exciton energies for three different zigzag SWCNTs—(10,0), (13,0), and (17,0)—against the experimental energies (Weisman and Bachilo 2003). The theoretical exciton energies are calculated using four different $t$ at 1.8, 1.9, 2.0, and 2.4 eV, and the results are shown in Table 2.1. Here and henceforth, we use $E_{11}$ and $E_{b1}$ to denote the absolute energy and the binding energy of the lowest longitudinal exciton Ex1, which consists predominantly of excitations from the highest valence band to the lowest conduction band. Similarly, $E_{22}$ and $E_{b2}$ are the energy and binding energy of the second lowest exciton Ex2, which is dominated by excitations from the second highest valence band to the second lowest conduction band.

**TABLE 2.1**
Calculated and Experimental $E_{11}$ and $E_{22}$ for Three Zigzag S-SWCNTs

| | | $E_{11}$ (eV) | | $E_{22}$ (eV) | |
|---|---|---|---|---|---|
| **(*n*,*m*)** | ***t* (eV)** | **SCI** | **Expt.** | **SCI** | **Expt.** |
| (10,0) | 1.8 | 1.10 | 1.07 | 1.97 | 2.31 |
| | 1.9 | 1.14 | | 2.05 | |
| | 2.0 | 1.18 | | 2.13 | |
| | 2.4 | 1.33 | | 2.45 | |
| (13,0) | 1.8 | 0.90 | 0.90 | 1.59 | 1.83 |
| | 1.9 | 0.93 | | 1.65 | |
| | 2.0 | 0.96 | | 1.71 | |
| | 2.4 | 1.08 | | 1.96 | |
| (17,0) | 1.8 | 0.73 | 0.80 | 1.24 | 1.26 |
| | 1.9 | 0.75 | | 1.28 | |
| | 2.0 | 0.77 | | 1.32 | |
| | 2.4 | 0.87 | | 1.50 | |

*Source:* Wang, Z. et al., *Phys. Rev. B* 74, 195406, 2006. With permission.

The computational results of Table 2.1 indicate that $t = 2.4$ eV—the standard hopping parameter for PCPs used by Zhao and Mazumdar (2004) initially for S-SWCNTs—is too large and considerably better fits are obtained with $t = 1.8$–$2.0$ eV. The choice of $t = 2.0$ eV has been used for all S-SWCNTs in the rest of this chapter. The fits in Table 2.1 improve with increasing nanotube diameter, implying that, strictly speaking, the hopping integral is diameter dependent. No attempt to further fine tune the parameters is made because that would necessarily lead to loss of simplicity and generality.

We have used the open boundary condition (OBC) in our calculations (Zhao and Mazumdar 2004, 2007; Zhao et al. 2006; Wang et al. 2006; Wang, Zhao, et al. 2007) because this enables precise determinations of transition dipole moments. With the OBC, surface states due to dangling bonds at the nanotube ends appear in the HF band structure and are discarded prior to the SCI stage of our calculations. In general, chiral S-SWCNTs have large unit cells. The number of unit cells we retain depends on the size of the unit cell and on the convergence behavior of $E_{11}$. The procedure involved calculating the standard n.n. tight-binding (TB) band structure with periodic boundary condition (PBC), and then comparing the PBC $E_{11}$ with that obtained using OBC with a small number of unit cells. The number of unit cells in the OBC calculation is now progressively increased until the difference in the computed $E_{11}$ between OBC and PBC is less than 0.004 eV (worst case).

It is with this system size that the SCI calculations are now performed using OBC. Thus, for example, our calculations for (7,0), (6,4), (7,5), and (8,4) SWCNTs are for 70, 16, 5, and 22 unit cells, respectively, and contain 1,960, 2,432, 2,180, and 2,464 carbon atoms, respectively. Because energy convergences are faster in the calculations with nonzero e–e interactions than the calculations in the n.n. TB limit, we are confident that this procedure gives accurate results. We retain an active space of 100 valence and conduction band states each in the SCI calculations. Stringent convergence tests involving gradual increase in the size of the active space indicate that the computational errors due to the energy cutoff are less than 0.005 eV (worst case).

### 2.5.2 Nanotube Transverse Excitons

Early theoretical investigation of the optical absorption in S-SWCNTs was by Ajiki and Ando (1994). They showed that when light is polarized perpendicular to the S-SWCNT axis, because of the induced local charges on S-SWCNTs, the absorption is suppressed "almost completely"; this is called the "depolarization effect." This conclusion was not challenged until about ten years later. Based on the previous experience with PCPs, Zhao and Mazumdar (2004) predicted that transverse optical absorptions, with smaller amplitude but far from invisible, will occur at energy that is higher than the middle of the lowest two longitudinal absorptions. Significant blue shift of the transverse absorption has also been found more recently within an effective mass approximation theory (Uryu and Ando 2006).

Experimentally, excited states coupled to the ground state by the transverse component of the dipole operator have recently been detected by polarized photoluminescence excitation (PLE) spectroscopy. Miyauchi, Oba, and Maruyama (2006) have determined the PL spectra for four chiral SWCNTs with diameters $d$ = 0.75–0.9 nm. In all cases, the allowed transverse optical absorption is close to $E_{22}$. Lefebvre and Finnie (2007) have detected transverse absorptions close to $E_{22}$ in 25 S-SWCNTs with even larger diameters; they were also able to demonstrate the same "family behavior" in transverse absorption energies that had been noted previously with longitudinal absorptions (Bachilo et al. 2002).

#### *2.5.2.1 One-Electron Limit*

We begin the discussion with the $U = V_{ij} = 0$ TB limit of Equation (2.1) for both PCPs and S-SWCNTs, in order to demonstrate the similarities between them. We also investigate the consequence of including next nearest neighbor (n.n.n.) hopping $t_2$ to determine the effect of broken CCS. In Figure 2.7(a)–(c), we show the TB energy structure for (10,0), (8,8), and (6,4) SWCNTs, examples of zigzag, armchair, and chiral SWCNTs, respectively. We label the conduction bands with letter "c" and valence bands with letter “v,” followed by their orders counted from Fermi energy. Figure 2.7(d) illustrates schematically the lowest two conduction and the highest two valence bands. According to TB theory, a (*n,m*) SWCNT is metallic if $|n-m| = 3j$, where $j$ is an integer. This is shown in Figure 2.7(b) for (8,8) tube, where *c*1 and *v*1 bands cross. For all other cases, SWCNTs are semiconductors.

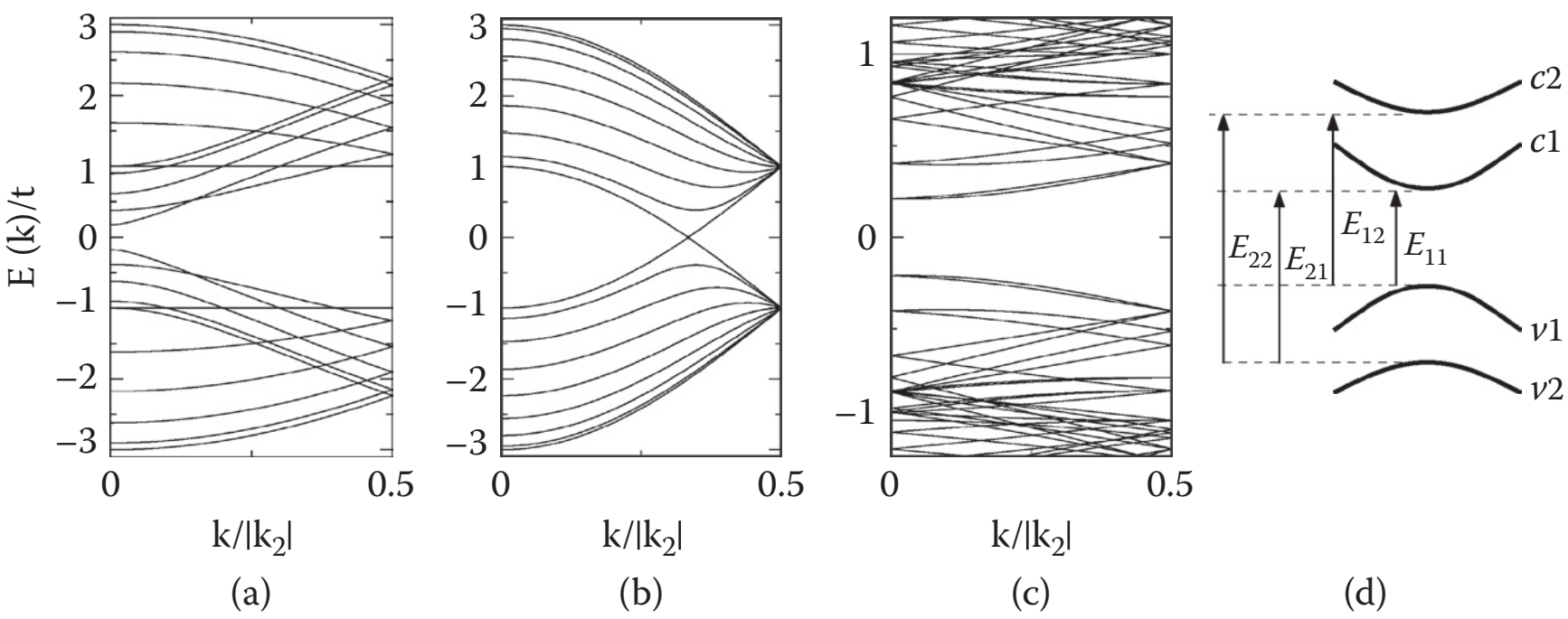

**FIGURE 2.7**
Tight binding energy structures for (a) (10,0) zigzag NT, (b) (8,8) armchair NT, and (c) (6,4) chiral NT; (d) shows the labeling of conduction and valence bands of SWCNTs.

We denote the TB energies as $E_{12}^{TB}$ and $E_{21}^{TB}$, corresponding to the transverse excitations $\psi_{v1\to c2}$ and $\psi_{v2\to c1}$, respectively (see Figure 2.7d). These two transitions are degenerate when $t_2 = 0$ and occur exactly at the center of the two longitudinal transitions at $E_{11}^{TB}$ and $E_{22}^{TB}$. This is shown using a dashed line in Figure 2.8(a) for the (6,5) S-SWCNT. The n.n.-only TB band structure of PPV is similar (see discussions in Section 2.4.1), although the nomenclature is different. The one-electron absorption spectrum for PPV is shown in Figure 2.8(b) for $t = 2.4$ eV, which is appropriate for planar $\pi$-conjugated systems. The oscillator strength of the central transverse absorption peak, relative to those for the longitudinal transitions, is much larger in the S-SWCNT than in PPV. This is a reflection of the larger electron–hole separation that is possible in the transverse direction in a S-SWCNT with $d \sim 1$ nm, as compared to PPV.

The degeneracy between $E_{12}^{TB}$ and $E_{21}^{TB}$ ($E_{dl^*}^{TB}$ and $E_{ld^*}^{TB}$ in PPV) is lost if CCS is broken by including nonzero $t_2$. The solid lines in Figure 2.8 show the effect of $t_2 = 0.6$ eV on the absorption spectra of the (6,5) S-SWCNT and PPV; we will argue later that this is the largest possible n.n.n. hopping between $\pi$-orbitals. The splitting between the transverse transitions is much smaller in the (6,5) S-SWCNT than in PPV. We have found this to be true for all four S-SWCNTs that we have studied. We do not show results of including a third-neighbor hopping because this does not contribute any further to the splitting of the transverse states.

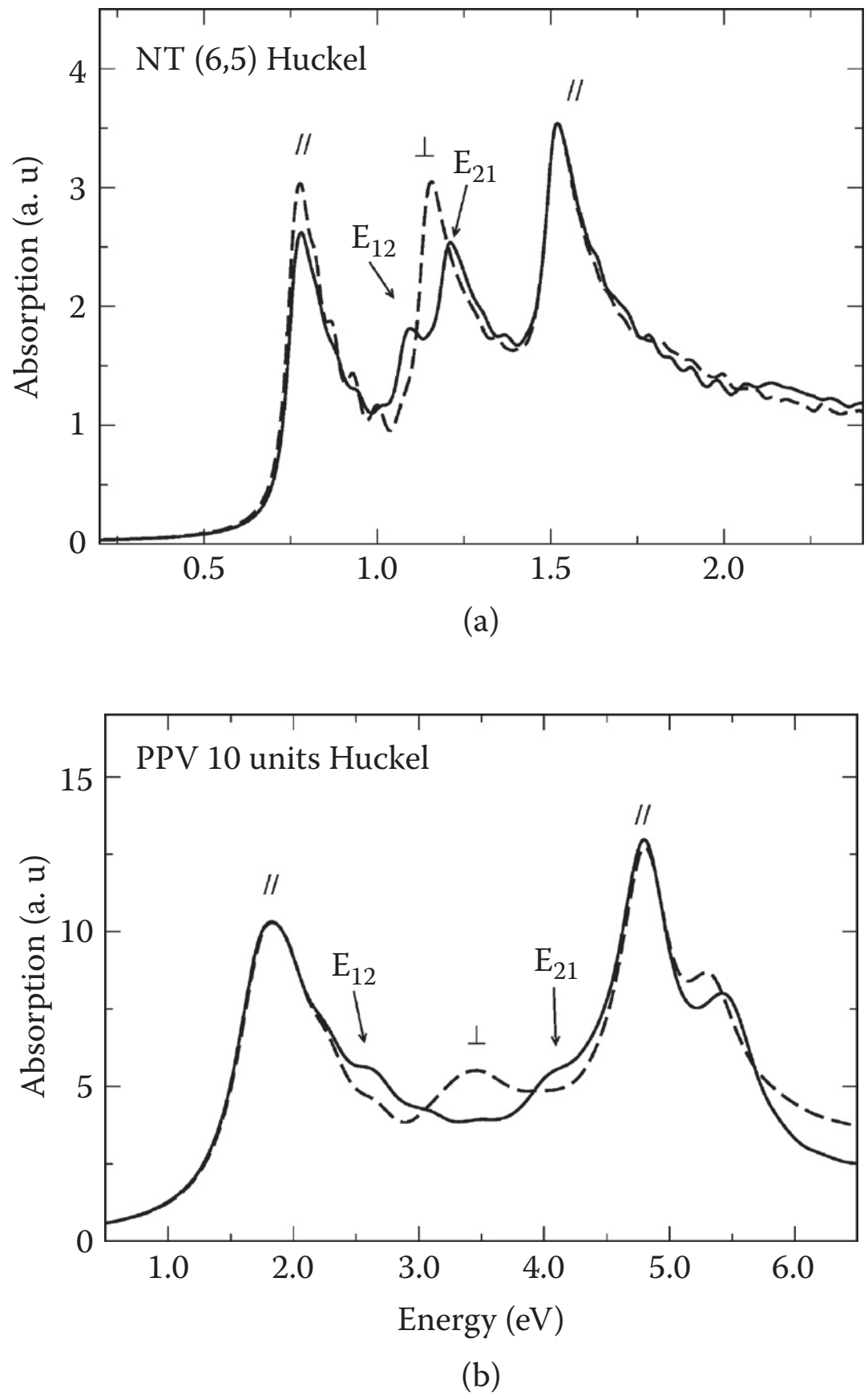

**FIGURE 2.8**
Calculated optical absorption spectra of (a) a (6,5) SWCNT, and (b) PPV within the tight binding model for $t_2 = 0$ (dashed line) and $t_2 = 0.6$ eV (solid line). The longitudinal (//) and perpendicular (⊥) components of the optical absorption are indicated in each case. (From Wang, Zhao, et al., *Phys. Rev. B*, 76, 115431, 2007. With permission.)

### *2.5.2.2 Nonzero Coulomb Interaction*

The matrix element $\langle \psi_{v1\to c2} | H_{ee} | \psi_{v2\to c1} \rangle$ is nonzero, and exactly as in PPV, the eigenstates of the Hamiltonian are now odd and even superpositions of these basis functions. In Figure 2.9, we have shown our calculated optical absorption spectra within Equation (2.1) for $t = 2.0$ eV, $t_2 = 0$, for all four S-SWCNTs investigated by Miyauchi et al. (2006). In all cases, the optically allowed transverse exciton is seen to occur very close to the Ex2, as observed experimentally. The relative oscillator strength of the transverse exciton is now considerably weaker than those of Ex1 and Ex2, in contrast to the calculations within TB theory as in Figure 2.8(a).

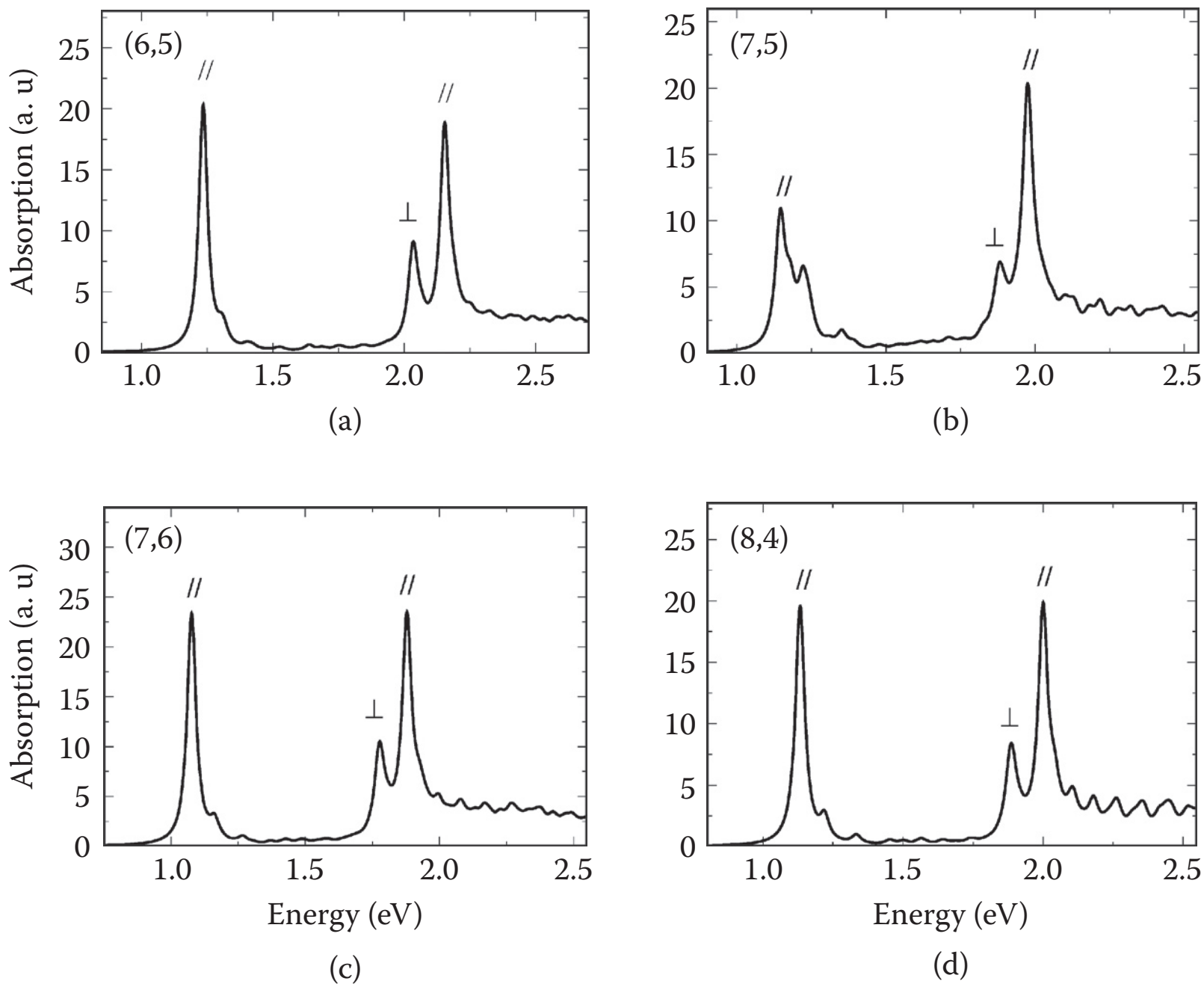

**FIGURE 2.9**
Calculated optical absorption spectra within the PPP model, with $t = 2.0$ eV, $t_2 = 0$, for four chiral SWCNTs. The longitudinal and transverse components of the absorption are indicated in each case. (From Wang, Zhao, et al., *Phys. Rev. B*, 76, 115431, 2007. With permission.)

We emphasize that the coupling between $\psi_{v1\to c2}$ and $\psi_{v2\to c1}$ is independent of the boundary condition (periodic versus open) along the longitudinal direction: The calculations for PPV in Rice and Gartstein (1994) and Gartstein, Rice, and Conwell (1995), for example, employed PBC and found results similar to those in our work. In the present case, we have repeated our calculations of all energies and wavefunctions, but not transition dipole couplings—which are difficult to define with PBC within TB models (Kuwata-Gonokami et al. 1994)—for all four S-SWCNTs and also with PBC. In every case, we have confirmed the splitting of the transverse wavefunctions into $\Psi_O$ and $\Psi_E$ from wavefunction analysis. We have confirmed that the energy differences between the odd and even superpositions are the same with the two boundary conditions for the number of unit cells used in the calculation. Our parameterization of the $V_{ij}$ involves a dielectric constant (Zhao and Mazumdar 2004; Wang et al. 2006). The energy splitting between the odd and even superpositions will occur for any finite dielectric constant, and only the magnitude of the splitting depends on the value of the dielectric constant.

### *2.5.2.3 Transverse Exciton and Its Binding Energy*

We compare experimental and calculated transverse optical absorptions of the S-SWCNTs in Figure 2.10. The peak heights of the calculated absorption spectra in Figure 2.10 have been adjusted to match those of the experimental spectra. The experimental absorption spectra show two peaks with nearly the same separation in all four cases: ~0.1 eV. Independent of which of these two peaks correspond to the true electronic energy of $\Psi_E$, it is clear that the error in our calculated energies is small: ≤0.1 eV. Within the SCI approximation, the lower edge of the continuum band is the HF threshold. We have indicated the HF thresholds for the transverse states in Figure 2.10. The binding energies of the transverse excitons, taken as the difference between the HF threshold and the exciton energy, are ~0.15 eV for all four SWCNTs. We will see that this is about one third of that of longitudinal Ex1. From a different perspective, Miyauchi et al. (2006) have also arrived at the conclusion that the binding energy of the transverse exciton is small. A similar conclusion for the transverse exciton in PPV was reached from photoconductivity studies (Köhler et al. 1998).

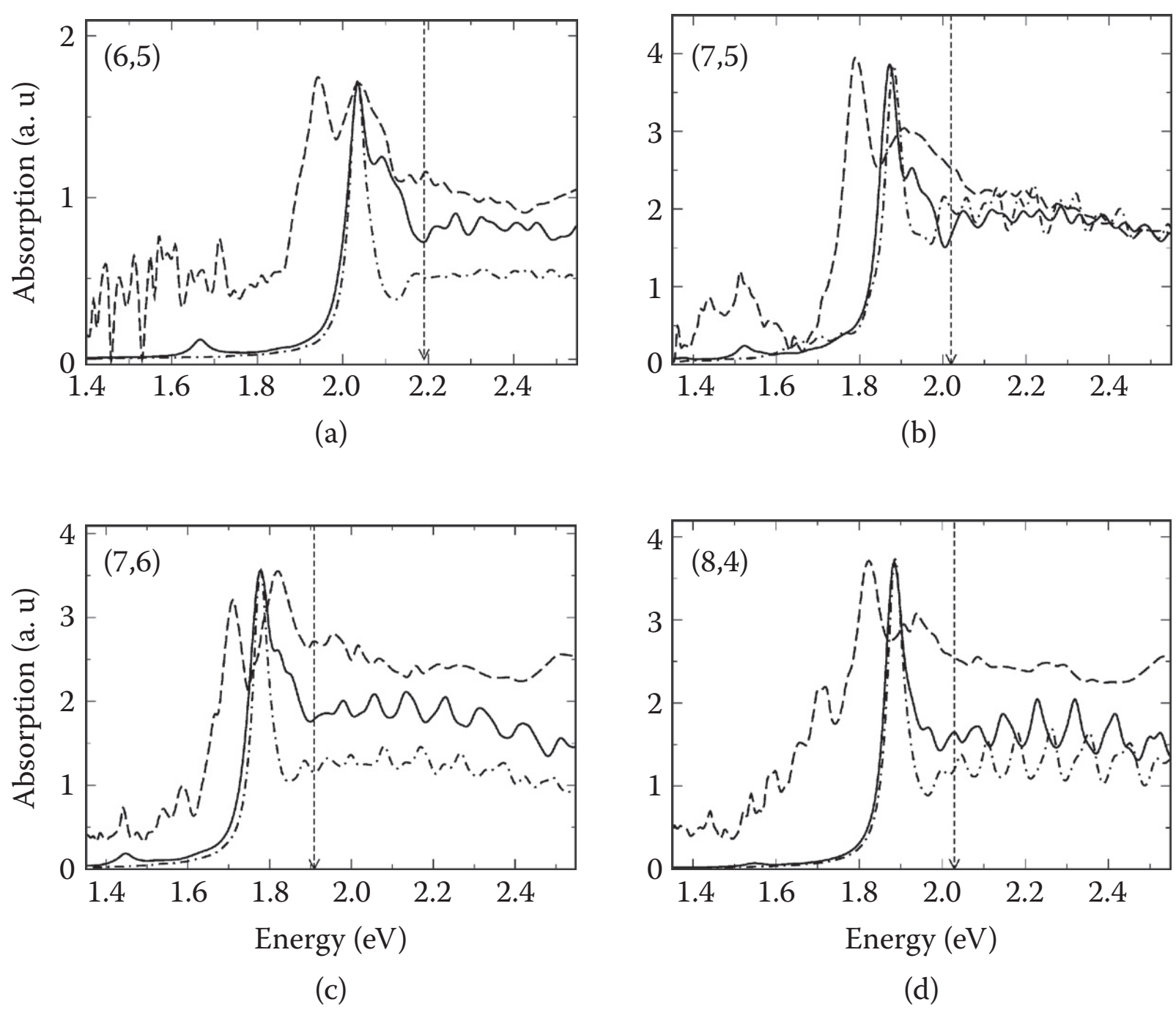

**FIGURE 2.10**
Comparison of experimental (dashed curves) and calculated transverse components of optical absorptions in four S-SWCNTs for $t_2 = 0$ (dot-dashed curves) and $t_2 = 0.6$ eV (solid curves). The vertical lines correspond to the Hartree–Fock threshold. (From Wang, Zhao, et al., *Phys. Rev. B*, 76, 115431, 2007. With permission.)

#### *2.5.2.4 Splitting of the Allowed Transverse Optical Absorption*

We now discuss the energy splitting of ~0.1 eV between the peaks in the experimental absorption spectra. Miyauchi et al. (2006) ascribe this to broken CCS—that is, nondegenerate $E_{12}^{TB}$ and $E_{21}^{TB}$ even within $H_{1e}$. Our calculated absorption spectra in Figure 2.10 for $t_2$ as large as 0.6 eV (nearly one third of $t$), however, fail to reproduce this splitting. This is to be anticipated from the $H_{ee} = 0$ absorption spectrum of Figure 2.8(a) because any splitting due to broken CCS, a one-electron effect, can only be smaller for $H_{ee} \neq 0$. At the same time, $t_2 = 0.6$ eV should be considered as the upper limit for the n.n.n. electron hopping based on experiments in PPV, as we explain later.

The experimental absorption spectra of PPV derivatives (see Figure 2.4a) resemble qualitatively the tight-binding absorption spectrum of Figure 2.8(b) for $t_2 \neq 0$, with, however, an overall blue shift due to e–e interactions. The spectra contain strong absorption bands at ~2.2–2.4 eV and 6 eV, and weaker features at 3.7 and 4.7 eV, respectively. The lowest and highest absorption bands are polarized predominantly along the polymer chain axis, while the 4.7 eV band is polarized perpendicular to the chain axis (Chandross et al. 1997; Comoretto et al. 1998, 2000; Miller et al. 1999). It is then tempting, based on Figure 2.8(b), to ascribe the origin of the 3.7-eV band in Figure 2.4(a) to broken CCS, in which case it ought to have the same polarization as the 4.7-eV absorption band. Repeated experiments have found, however, that the absorption band at 3.7 eV is polarized predominantly along the polymer chain axis. Theoretical calculations within the PPP model with $t_2 = 0$ have reproduced the longitudinal polarization of the 3.7-eV band (Chandross et al. 1997), which is ascribed to the second lowest longitudinal exciton in PPV derivatives. We have confirmed that inclusion of $t_2 = 0.6$ eV within the same PPP model calculation renders the polarization of the 3.7-eV absorption band perpendicular to the polymer chain direction, in contradiction of experiments. The n.n.n. hopping in PPV is therefore certainly smaller than 0.6 eV.

Curvature of SWCNTs implies an even smaller value of $t_2$ in SWCNTs (Wang et al. 2006). We are consequently unable to give a satisfactory explanation of the splitting in the transverse absorption in S-SWCNTs. Ando (1997) has ascribed the splitting, within the $\boldsymbol{k} \times \boldsymbol{p}$ scheme, to a higher energy transverse exciton. We have not found any higher energy transverse exciton whose oscillator strength can explain the observed absorption spectra. It is conceivable that the second peak in the experimental absorption spectra in Figure 2.10 corresponds to the threshold of the transverse continuum band (see, in particular, the spectra in Figure 2.10b

AU: pls add to ref list

and 2.10d). Further experimental work is therefore warranted. It is also possible that the energy splitting is due to higher order correlation effects neglected in SCI or to intertube interactions.

### 2.5.3 Longitudinal Excitons in S-SWCNTs

In the previous section, we have shown the important role of e–e interactions in S-SWCNTs. We have also demonstrated the excellent agreement between our theoretical results for transverse excitons in S-SWCNTs and recent experiments. In this section, we present our theoretical results for the linear longitudinal optical excitations in S-SWCNTs, which match those of many experiments.

#### *2.5.3.1 Dark Excitons and Electronic Structure with Many-Body Coulomb Interactions*

Within n.n. TB theory, multiple pairs of valence and conduction bands exist in S-SWCNTs. The conduction and valence bands close to Fermi energy are exactly doubly degenerate for achiral nanotubes and are almost degenerate for chiral ones. This is a consequence of cylindrical symmetry of quasi one-dimensional geometry of SWCNTs. Thus, corresponding to each pair of valence and conduction bands, four degenerate excitations occur. Consider now the four degenerate lowest single-particle excitations in S-SWCNTs, $\chi_{a\to a'}$, $\chi_{a\to b'}$, $\chi_{b\to a'}$, and $\chi_{b\to b'}$, where $a$, $b$ ($a'$, $b'$) are the highest occupied (lowest unoccupied) TB levels. The two excitations $\chi_{a\to a'}$ and $\chi_{b\to b'}$ are optically allowed, and the excitations $\chi_{a\to b'}$ and $\chi_{b\to a'}$ are forbidden. As in the case of the transverse excitations, nonzero $H_{ee}$ splits these degenerate one-electron excitations into $\chi_{a\to a'} \pm \chi_{b\to b'}$ and $\chi_{a\to b'} \pm \chi_{b\to a'}$.

For the pair of dipole-allowed TB excitations, the odd superposition is optically inactive, or dark; the even one is optically active, or bright. Both the odd and even superpositions of dipole-forbidden excitations are dark. As a consequence, $H_{ee}$ splits the four degenerate TB excitations into one optical exciton and three dark excitons. For repulsive $H_{ee}$, the optical exciton is always higher in energy than the dark ones. The longitudinal energy spectra of SWCNTs is similar to that of $t$-$(CH)_x$ and PDAs, where dark excitons also occur below the optical exciton as a consequence of e–e interaction (Hudson et al. 1982; Soos et al. 1992). (One fundamental difference, however, is that the dark states in $t$-$(CH)_x$ and PDAs are two photon allowed; this is not so for the S-SWCNTs.) PL is weak in these polymers (Burroughes et al. 1990; Colaneri et al. 1990; Swanson et al. 1991) because the optically excited state decays in ultrafast times to the low-energy dark exciton, radiative transition from which to the ground state cannot occur. The occurrence of dark excitons below the optical exciton suggests that the low quantum efficiency of the PL of S-SWCNTs (O'Connell et al. 2002; Lebedkin et al. 2003; Wang et al. 2004) may be intrinsic.

The longitudinal dark excitons in S-SWCNTs have attracted enormous interest (Sheng et al. 2005; Seferyan et al. 2006; Satishkumar et al. 2006; Zaric et al. 2006; Shaver et al. 2007; Zhu et al. 2007; Jones et al. 2007; Berger et al. 2007; Mortimer and Nicholas 2007; Kiowski et al. 2007) ever since their prediction in 2004 by Zhao and Mazumdar. Magnetic brightening of the dark excitons of S-SWCNTs has been observed in the experimental PL spectra (Zaric et al. 2004, 2006; Shaver et al. 2007), which gives direct evidence of their existence. Other experiments that have verified the existence of dark excitons include transient grating measurement (Seferyan et al. 2006), Raman scattering (Satishkumar et al. 2006), and temperature-dependent PL (Berger et al. 2007; Mortimer and Nicholas 2007; Scholes et al. 2007; Kiowski et al. 2007), as well as PL decay of large number of S-SWCNTs using time-correlated single photon counting (Jones et al. 2007).

The splitting of the bright and dark excitons has been claimed to be only a few millielectronvolts by Mortimer and Nicholas (2007) and as large as 0.05–0.1 eV by Scholes et al. (2007) and 0.1–0.14 eV by Kiowski et al. (2007). The theoretical estimation of bright–dark exciton gap is difficult from our calculations. In principle, the energy gap can be obtained from calculations of the energetics of very long nanotubes. Because of the small size of the bright–dark energy gap, however, convergence in this quantity is very hard to reach, even well after convergence in the absolute energies $E_{11}$ and $E_{22}$ of the bright excitons have been reached. Should this energy splitting really be only a few millielectronvolts—smaller than the thermal fluctuation at room temperature—our original explanation of the low quantum efficiency of emission of S-SWCNTs becomes questionable. The resolution of this question needs attention.

#### 2.5.3.2 Energy Manifolds

One-electron TB calculations indicate that dipole-allowed longitudinal excitations occur only between valence and conduction bands placed symmetrically about the chemical potential (see Figure 2.7d). Within a total energy scheme, it is then possible to define energy manifolds $n = 1, 2, 3, \ldots$, etc., where the $n = 1$ manifold consists of excitations of the type $E_{11}$, the $n = 2$ manifold consists of excitations of the type $E_{22}$, and so on. The different manifolds are clearly independent of one another in the noninteracting limit. Interestingly, our calculations indicate that relatively weak mixing occurs between the different $n$ excitations, even with nonzero $H_{ee}$, and the classification into manifolds continues to be meaningful. The energy spectra of S-SWCNTs then consist of a series of total energy manifolds, whose energies increase with their index $n$. Within each energy manifold, the single optical exciton and several dark excitons occur, as well as a continuum band separated from the optical exciton by a characteristic exciton binding energy. This is shown in Figure 2.11, where we compare schematically the electronic structures of a PPV derivative and an S-SWCNT.

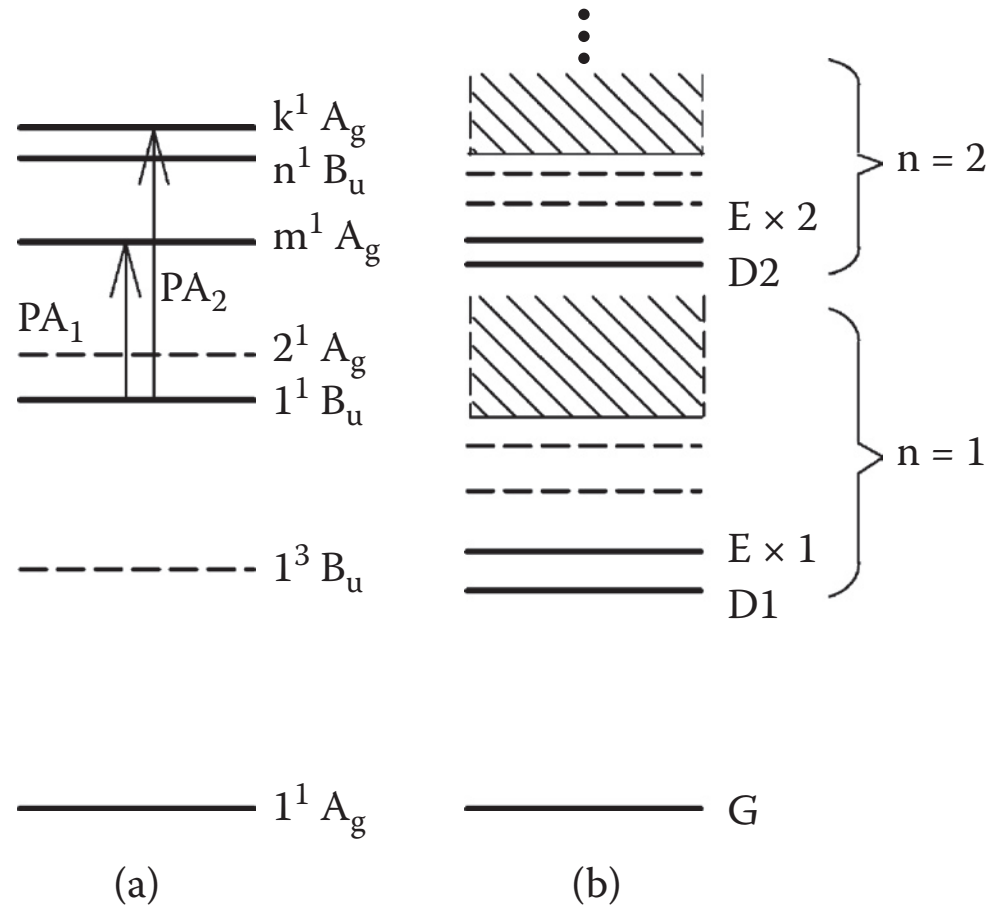

**FIGURE 2.11**
Schematics of the excitonic electronic structures of (a) a light-emissive $\pi$-conjugated polymer, and (b) an S-SWCNT. In (a), the lowest triplet exciton $1^3B_u$ occurs below the lowest singlet exciton $1^1B_u$. The lowest two-photon state $2^1A_g$ is composed of two triplets and plays a weak role in nonlinear absorption. Transient PA is from the $1^1B_u$ to the $m^1A_g$ two-photon exciton, which occurs below the continuum band threshold state $n^1B_u$, and to a high energy $k^1A_g$ state that occurs deep inside the continuum band. In (b), $n = 1$ and $n = 2$ energy manifolds for S-SWCNT are shown. Ex*n* and D*n* are dipole-allowed and -forbidden excitons, respectively. Shaded areas indicate continuum band for each manifold. (Note that all levels of the $n = 2$ manifold should be buried in the $n = 1$ continuum band; however, for clarity, the $n = 1$ continuum band ends below D2.) (From Zhao, H. et al., *Phys. Rev. B*, 73, 075403, 2006. With permission.)

Nonlinear spectroscopic measurements in S-SWCNTs have demonstrated that additional optically relevant states occur within each *n*. Specifically, these include the dominant two-photon states, equivalent to the $m^1A_g$ that we have already discussed in the context of PCPs. We discuss this further in Section 2.5.4.

#### *2.5.3.3 Longitudinal Exciton Energies and Their Binding Energies*

We present our calculated longitudinal energy spectra of 29 different S-SWCNTs within Equation (2.1). The diameters of the S-SWCNTs we consider range from 0.56 to 1.51 nm. For each S-SWCNT, we calculate the absolute energies $E_{11}$ and $E_{22}$ and their binding energies $E_{b1}$ and $E_{b2}$. We compare all theoretical quantities to experimentally determined ones (Bachilo et al. 2002; Weisman and Bachilo 2003; Fantini et al. 2004; Wang et al. 2005; Dukovic et al. 2005; Maultzsch et al. 2005; Ma et al. 2005; Zhao et al. 2006). The large number of S-SWCNTs that could be considered allows us to investigate family relationships that have been demonstrated by experimentalists (Bachilo et al. 2002; Reich, Thomsen, and Robertson 2005). We find excellent agreement between the theory and experiments.

**TABLE 2.2**

Comparison of Calculated and Experimental/Empirical $n = 1$ and $n = 2$ Exciton Energies and Binding Energies

| | | $E_{11}$ (eV) | | $E_{22}$ (eV) | | $E_{b1}$ (eV) | | $E_{b2}$ (eV) |
|---|---|---|---|---|---|---|---|---|
| **(*n*,*m*)** | ***d*** **(nm)** | **SCI** | **Expt.**[a] | **SCI** | **Expt.** | **SCI** | **Expt.**[c] | **SCI** |
| (7,0) | 0.56 | 1.58 | (1.29) | 2.92 | (3.14) | 0.56 | (0.61) | 0.79 |
| (6,2) | 0.57 | 1.55 | (1.39) | 2.82 | (2.96) | 0.55 | (0.59) | 0.72 |
| (8,0) | 0.64 | 1.44 | (1.60) | 2.38 | (1.88) | 0.56 | (0.54) | 0.57 |
| (7,2) | 0.65 | 1.41 | (1.55) | 2.36 | (1.98) | 0.54 | (0.52) | 0.56 |
| (8,1) | 0.68 | 1.34 | (1.19) | 2.45 | (2.63) | 0.48 | (0.50) | 0.65 |
| (6,4) | 0.69 | 1.33 | 1.42 | 2.27 | 2.13[a,b] | 0.50 | (0.49) | 0.56 |
| (6,5) | 0.76 | 1.24 | 1.27 | 2.15 | 2.19[a,b] | 0.45 | 0.43 | 0.54 |
| (9,1) | 0.76 | 1.24 | 1.36 | 2.08 | 1.79[a,b] | 0.47 | (0.45) | 0.51 |
| (8,3) | 0.78 | 1.21 | 1.30 | 2.05 | 1.87[a] | 0.45 | 0.42 | 0.50 |
| (10,0) | 0.79 | 1.18 | (1.07) | 2.13 | 2.26[b] | 0.42 | (0.43) | 0.57 |
| (9,2) | 0.81 | 1.17 | (1.09) | 2.10 | 2.24[b] | 0.42 | (0.42) | 0.55 |
| (7,5) | 0.83 | 1.15 | 1.21 | 1.97 | 1.93[a,b] | 0.43 | 0.39 | 0.49 |
| (8,4) | 0.84 | 1.13 | 1.11 | 2.00 | 2.11[a,b] | 0.41 | (0.40) | 0.51 |
| (11,0) | 0.87 | 1.11 | (1.20) | 1.86 | (1.67) | 0.42 | (0.39) | 0.46 |
| (10,2) | 0.88 | 1.09 | (1.18) | 1.84 | 1.68[b] | 0.40 | 0.34 | 0.45 |
| (7,6) | 0.90 | 1.08 | 1.11 | 1.88 | 1.92[a,b] | 0.39 | 0.35 | 0.47 |
| (9,4) | 0.92 | 1.06 | 1.13 | 1.81 | 1.72,[a] 2.03[b] | 0.39 | 0.34 | 0.44 |
| (11,1) | 0.92 | 1.05 | (0.98) | 1.89 | (2.03), 1.72[b] | 0.37 | (0.37) | 0.50 |
| (10,3) | 0.94 | 1.03 | 0.99 | 1.84 | 1.96[a,b] | 0.37 | (0.36) | 0.48 |
| (8,6) | 0.97 | 1.01 | 1.06 | 1.75 | 1.73[a,b] | 0.37 | 0.35 | 0.44 |
| (13,0) | 1.03 | 0.96 | (0.90) | 1.71 | (1.83) | 0.34 | (0.33) | 0.45 |
| (12,2) | 1.04 | 0.95 | 0.90 | 1.69 | 1.81[a] | 0.33 | (0.33) | 0.44 |
| (10,5) | 1.05 | 0.94 | 0.99 | 1.62 | 1.58[a,b] | 0.35 | (0.32) | 0.40 |
| (14,0) | 1.11 | 0.91 | (0.96) | 1.54 | (1.44) | 0.34 | (0.31) | 0.38 |
| (12,4) | 1.15 | 0.88 | 0.92 | 1.51 | 1.45[a] | 0.32 | 0.27 | 0.37 |
| (16,0) | 1.27 | 0.81 | (0.76) | 1.44 | (1.52) | 0.28 | (0.27) | 0.37 |
| (17,0) | 1.35 | 0.77 | (0.80) | 1.32 | (1.26) | 0.28 | (0.25) | 0.32 |
| (15,5) | 1.43 | 0.73 | (0.71) | 1.29 | (1.35) | 0.25 | (0.24) | 0.32 |
| (19,0) | 1.51 | 0.70 | (0.66) | 1.24 | (1.30) | 0.24 | (0.23) | 0.31 |

*Source:* Wang, Z. et al., *Phys. Rev. B* 74, 195406, 2006. With permission.
*Note:* The empirical exciton energies (Weisman and Bachilo 2003) and exciton binding energies (Dukovic et al. 2005) are in parentheses.

[a]From Bachilo et al. (2002).
[b]From Fantini et al. (2004).
[c]From Dukovic et al. (2005).

Our work demonstrates convincingly that the photophysics of S-SWCNTs and PCPs can be understood within the same general theoretical framework, albeit with different hopping integrals.

In Table 2.2 we have listed our calculated $E_{11}$, $E_{22}$, $E_{b1}$, and $E_{b2}$ for 29 S-SWCNTs. We compare each of these quantities to those obtained by experimental investigators (Bachilo et al. 2002; Weisman and Bachilo 2003; Fantini et al. 2004; Ma et al. 2005; Dukovic et al. 2005). Nearly half the exciton energies listed as experimental in Table 2.2 were obtained directly from spectrofluorometric measurements (Bachilo et al. 2002) or from resonant Raman spectroscopy (Fantini et al. 2004). The other half are empirical quantities arrived at by Weisman and Bachilo (2003). Using the experimental data in Bachilo et al. (2002), Weisman and Bachilo (2003) derived empirical equations for the exciton energies of nanotubes for which direct experimental information does not exist. The experimental $E_{11}$ and $E_{22}$ in Table 2.1 are obtained from these empirical equations. Dukovic et al. (2005) have also given an empirical equation for the binding energy of Ex1, which was also derived by fitting the set of $E_{b1}$ obtained from direct measurements. We make distinctions between the experimental and empirical data in Table 2.2, but in the following text we refer to both as experimental quantities.

In Figure 2.12, we have plotted the theoretical and experimental $E_{11}$ and $E_{22}$ against $1/d$ (where $d$ is the diameter of the tube), and in the inset we show the errors in our calculations, $\Delta E_{11}$ and $\Delta E_{22}$, defined as the calculated energies minus the experimental quantities, for $d > 0.75$ nm.

The spreads in the experimental $E_{11}$ and $E_{22}$ are systematically larger than the calculated quantities; however, the latter do capture the effects due to differences in chirality qualitatively. The sudden increases or decreases in the experimental $E_{22}$ between the nearest data points in the experimental plot of Figure 2.12 are reflected correctly in the theoretical plot in all cases, even though the magnitudes of these changes are larger in the experimental data set. We find excellent agreement between calculated and experimental $E_{11}$ for $d > 0.75$ nm, with $|\Delta E_{11}| < 0.1$ eV. The agreement for $d > 1$ nm is even better with $|\Delta E_{11}| < 0.05$ eV. The disagreements between calculated and experimental $E_{22}$ are larger, but even here the magnitude of the maximum error for $d > 0.75$ nm is within 0.2 eV, which is the C–C bond stretching frequency that can influence experimental estimation of exciton energies (Perebeinos et al. 2004; Perebeinos, Tersoff, and Avouris 2005).

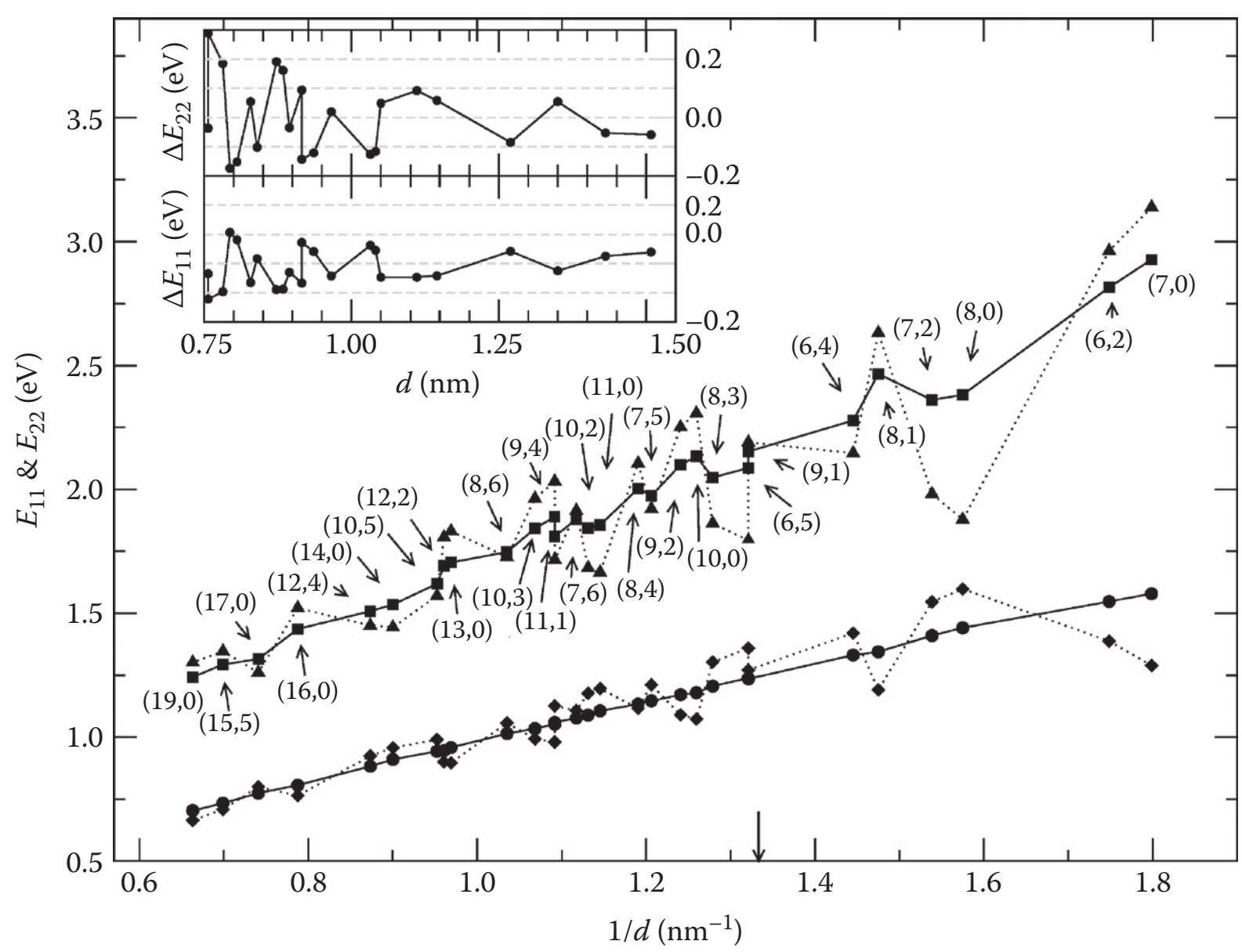

**FIGURE 2.12**
Calculated (solid line, circle and square symbols) versus experimental (dotted line, diamond and triangle symbols) $E_{11}$ and $E_{22}$ for 29 S-SWCNTs. The arrow against the x-axis corresponds to $d = 0.75$ nm. The inset shows errors $\Delta E_{ii}$ ($i = 1, 2$) in the calculations, defined as the calculated minus the experimental or empirical energies. (From Wang, Z. et al., *Phys. Rev. B,* 74, 195406, 2006. With permission.)

The relatively large disagreement between experimental and calculated energies for $d < 0.75$ nm is due to the breakdown of the $\pi$-electron approximation. The larger spread in the experimental exciton energies for $d <$ 0.75 nm can be due to the curvature effect and trigonal warping effect that are ignored in Equation (2.1). It is likely that in the case of $E_{22}$, greater precision will necessarily require inclusion of higher order CI. This is because SCI approximation works best for lower energy regions, where single excitations dominate the states. As the energy increases, higher order excitations contribute more to the wavefunctions; this requires higher order CI. Overall, however, the close agreement between theory and experiment in Figure 2.12 for $d > 0.75$ nm, the region where the $\pi$-electron model appears to be valid, is remarkable, given that a single semiempirical Hamiltonian with a single set of parameters is used to obtain the data.

The agreement between the calculated and the experimental exciton binding energies in Table 2.2 is even more striking than the fits to the absolute energies. The average error in the calculated exciton binding energies, when compared to the set of eight S-SWCNTs for which data are obtained directly from experiments, is only 0.039 eV. The average error for the complete set, including those S-SWCNTs for which only empirical data exist for the moment, is even less at 0.023 eV. The exciton binding energies depend weakly on chirality, in agreement with published experimental work (Dukovic et al. 2005). Indeed, we have found that $E_{b1}$ and $E_{b2}$ are both inversely proportional to the diameter $d$ and can be fit approximately by

$$E_{b1} \cong \frac{0.35}{d}\,\text{eV}, \tag{2.5}$$

$$E_{b2} \cong \frac{0.42}{d}\,\text{eV}, \tag{2.6}$$

Equation (2.5) is remarkably close to the empirical formula $E_{b1} \approx 0.34/d$ eV given in previous experimental work of Dukovic et al. (2005). For both binding energies, the fits of Equations (2.5) and (2.6) become better for larger diameter nanotubes.

The weak dependence of the exciton binding energy on chirality is very likely a cancellation effect: The effect of chirality on the exciton energy and the continuum band threshold energy for a given diameter are presumably similar. The calculated $E_{b1}$ and $E_{b2}$ are very close to those that we obtained earlier for the wide nanotubes with $t = 2.4$ eV (Zhao and Mazumdar 2004; Zhao et al. 2006), even as the calculated $E_{11}$ and $E_{22}$ are now quite different. The weak dependence of the exciton binding energy (but not the absolute exciton energy) on the magnitude of the hopping integral is simply a consequence of localization of the electrons at the large $U/t$ considered here. For such large Coulomb interaction, the dominant contribution to the exciton binding energy comes from the difference between $U$ and the n.n. intersite Coulomb interaction.

### 2.5.4 Ultrafast Spectroscopy of S-SWCNTs

In order to understand excited state absorptions we performed SCI calculations using the same Coulomb parameters. Zigzag S-SWCNTs possess inversion symmetry, and therefore nondegenerate eigenstates are once again classified as $A_g$ or $B_u$. Lack of inversion symmetry in chiral S-SWCNTs implies that their eigenstates are not strictly one- or two-photon states. Nevertheless, from explicit calculations of matrix elements of the dipole operator, we have found that even eigenstates of chiral S-SWCNTs are *predominantly* one-photon (with negligible two-photon cross-section) or *predominantly* two-photon (with very weak one-photon dipole coupling to the ground state) states. We shall therefore refer to eigenstates of chiral S-SWCNTs as $A_g$ and $B_u$, respectively.

PA in S-SWCNTs is due to excited state absorption from the $n = 1$ exciton states—from Ex1 as well as from D1—following rapid nonradiative decay of Ex1 to D1. As in the case of $\pi$-conjugated polymers (Chandross et al. 1999), we have evaluated all transition dipole couplings between the $n = 1$ exciton states (Ex1 and D1) and all higher energy excitations. Our computational results are the same for all zigzag nanotubes. These are modified somewhat for the chiral nanotubes (see later discussion), but the behavior of all chiral S-SWCNTs is again similar. In Figures 2.13 (a) and (b), we show the representative results for the zigzag (10,0) and the chiral (6,2) S-SWCNTs, respectively. The solid vertical lines in Figure 2.13(a) indicate the magnitudes of the normalized dipole couplings between Ex1 in the (10,0) NT with all higher energy excitations $e_j$, $\langle \text{Ex1}|\mu|e_j\rangle / \langle \text{Ex1}|\mu|G\rangle$, where $G$ is the ground state. The dotted vertical lines are the normalized transition dipole moments between the dark exciton D1 and the higher excited states, $\langle \text{D1}|\mu|e_j\rangle / \langle \text{Ex1}|\mu|G\rangle$. Both couplings are shown against the quantum numbers $j$ of the final state along the lower horizontal axis, while the energies of the states $j$ are indicated on the upper horizontal axis.

The reason why only two vertical lines appear in Figure 2.13(a) is that all other normalized dipole couplings are *invisible on the scale of the figure.*

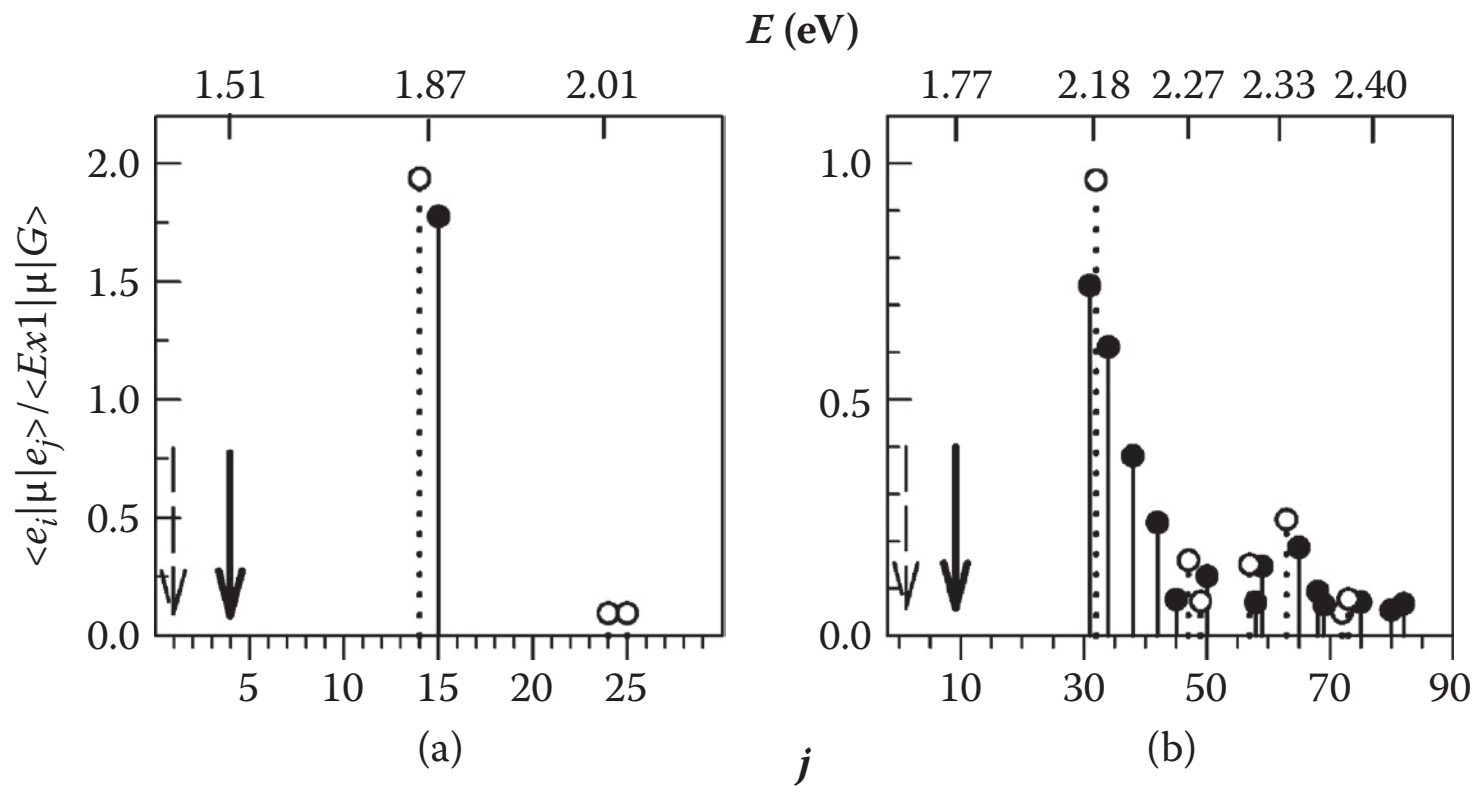

**FIGURE 2.13**
Normalized transition dipole moments between S-SWCNT exciton states Ex1 and D1 and all other excited states $e_j$, where $j$ is the quantum number of the state in the total space of single excitations from the HF ground state. The numbers along the upper horizontal axes are energies in electronvolts. Results shown are for (a) the (10,0), and (b) the (6,2) S-SWCNTs, respectively. Solid (dotted) lines correspond to $e_i$ = Ex1 (D1). The solid and dashed arrows denote the quantum numbers of Ex1 and D1, respectively. (From Zhao, H. et al., *Phys. Rev. B*, 73, 075403, 2006. With permission.)

A striking aspect of the results for the (10,0) zigzag S-SWCNT are then that *exactly as in the π-conjugated polymers, the optical exciton Ex1 is strongly dipole coupled to a single higher energy* $m^1A_g$ *state.* The dark exciton D1 is similarly strongly coupled to a single higher energy state (hereafter, the $m'^1A_g$). Furthermore, the dipole couplings between Ex1 and $m^1A_g$ (or D1 and $m'^1A_g$) are stronger than those between the ground state and the excitons, which is also true for the π-conjugated polymers (Chandross et al. 1999).

The situation in the chiral (6,2) S-SWCNT is slightly different, as shown in Figure 2.13(b). Both the Ex1 and D1 excitons are now strongly dipole coupled to several close-lying excited states, which form narrow "bands" of $m^1A_g$ and $m'^1A_g$ states. Similarly to the case of zigzag S-SWCNTs, these bands occur above the Ex1. Couplings between Ex1 and $m^1A_g$ (or D1 and $m'^1A_g$) are stronger than those between the ground state and the excitons, which is also true for the PCPs (Chandross et al. 1999).

Based on the preceding computational results, we then expect PA spectra of S-SWCNTs to be very similar to those of PCPs. Together with our experimental colleagues, we performed ultrafast pump–probe spectroscopy on films of isolated nanotubes with a diameter distribution around the mean diameter of ~0.8 nm (Zhao et al. 2006). The details of the sample preparation and the experimental setup can be found in the original reference. In Figure 2.5(c), we show the transient photomodulation spectrum of an S-SWCNT film. This figure should be compared to Figures 2.5(a) and (b). The similarity between the PAs in the two cases is striking. From the correlated dynamics of the transient photobleaching (PB) and PA bands, it has been concluded that PA originates from excitons in the $n = 1$ manifold (Korovyanko et al. 2004; Sheng et al. 2005). The lack of stimulated emission in the S-SWCNT photomodulation spectrum shows that whereas excitons in polymers are radiative, excitons in the S-SWCNTs are not. The dominance of nonradiative over radiative recombination in S-SWCNTs has been ascribed to a variety of effects, including (1) trapping of the excitation at defect sites (Wang et al. 2004), (2) strong electron–phonon coupling (Htoon et al. 2005), and (3) the occurrence of optically dark excitons below the allowed excitons (Zhao and Mazumdar 2004).

From the calculated results of Figure 2.13, a simple interpretation to $PA_1$ in Figure 2.5(c) emerges: $PA_1$ is a superposition of excited state absorptions from Ex1 and D1 to higher energy two-photon excitons. This raises the question whether $PA_2$ in the S-SWCNTs can be higher energy inter-subband absorptions from the $n = 1$ excitons to two-photon states that lie in the $n = 2$ (or even $n = 3$) manifolds. We have eliminated this possibility from explicit calculations: The transition dipole matrix elements between one-photon states in the $n = 1$ manifold and two-photon states within the higher $n$ manifolds are zero. Based on our experience with the PCPs, we therefore ascribe $PA_2$ in S-SWCNTs to many-body $k^1A_g$-like states (see Figure 2.6), which involve two-electron–two-hole excitations and multiple bands.

The broad nature of the $PA_1$ band in the S-SWCNTs arises from the inhomogeneous nature of the experimental sample, with SWCNT bundles that contain a distribution of S-SWCNTs with different diameters and exciton binding energies. If we assume that the peak in the $PA_1$ band corresponds to those S-SWCNTs that dominate nonlinear absorption, then the low energy of the peak in the $PA_1$ band in Figure 2.5(c) suggests that PA is dominated by the widest S-SWCNTs in our sample. The common origin of $PA_1$ and $PA_2$ then suggests that the peak in the $PA_2$ band at ~0.7 eV is also due to the widest S-SWCNTs, with $PA_2$ due to narrower S-SWCNTs occurring at even higher energies. Hence, the energy region 0.2–0.55 eV in Figure 2.5(c) must correspond only to $PA_1$ excitations. Based on the similarities in the energy spectra of the S-SWCNTs and the PCPs in Figure 2.11, we can therefore construct the vertical dashed line in Figure 2.5(c), which identifies the threshold of the continuum band for the widest S-SWCNTs in the film.

## 2.6 Conclusions and Future Work

We have, in the present work, focused on one specific aspect of the PCPs and S-SWCNTs: their related electronic structures and similar behavior under photoexcitation. As we have shown from detailed calculations, the exciton behavior in these systems can be understood quantitatively within the $\pi$-electron model. In the case of PPV, although the Coulomb parameters are obtained by fitting the linear absorption, no further modification of the parameters is done while calculating the nonlinear absorption spectra. Similarly, in the context of S-SWCNTs, we have demonstrated that the same model Hamiltonian with the same one-electron hopping and Coulomb interactions can reproduce the experimental energies and absorption spectra of longitudinal and transverse optical excitations in S-SWCNTs with diameters greater than 0.75 nm with considerable precision (errors ≤ 0.05–0.1 eV).

In cases where the experimental binding energies of Ex1 are known, the calculated quantities are uniformly very close (Wang et al. 2006). It has been suggested that the true single tube binding energies are considerably larger than the 0.3–0.4 eV that are found experimentally (Dukovic et al. 2005; Maultzsch et al. 2005; Zhao et al. 2006) for S-SWCNTs with diameters ~0.75–1 nm, and the experimental quantities reflect strong screening of e–e interactions by the environment. The close agreements between our theoretical single tube calculations and experiments suggest, however, that any such environmental effect on the exciton binding energy is small.

In our work here we have not discussed the effects of external electric or magnetic fields on the photophysics of PCPs and S-SWCNTs. Electroabsorption has been widely applied for many years now for investigating the electronic structures of PCPs (Sebastian and Weiser 1981). The theory of electroabsorption in PCPs is also well developed (Guo et al. 1993). Specifically, the same $m^1A_g$ and $n^1B_u$ states are observed in modulation spectroscopy. Theoretical work on electroabsorption has been extended to the S-SWCNTs recently (Perebeinos and Avouris 2007; Zhao and Mazumdar 2007). It has been pointed out that this technique offers the most straightforward approach to measure the binding energy of the $n = 2$ exciton (Zhao and Mazumdar 2007). Experimentally, complete separation of semiconducting and metallic nanotubes, which would be required for performing electroabsorption on the semiconducting samples, is still not feasible. We have already mentioned magnetic brightening of dark excitons (Zaric et al. 2004, 2006; Shaver et al. 2007). Recent experimental work on the energy gap between the bright and dark excitons (Scholes et al. 2007; Kiowski et al. 2007) suggests that there might be a need to revisit the theory of magnetic brightening.

Other directions for future work include:

Interchain effects on PCP photophysics. Our work on the photophysics of PCPs is valid only for single chains. Experimentally, single-chain behavior is found in dilute solutions and a few PPV derivatives with specific side groups. In most cases, thin films of PCPs exhibit very different behavior (Schwartz 2003; Arkhipov and Bässler 2004; Rothberg 2007), and various interchain species appear to dominate the photophysics. It has also been suggested that branching of photoexcitations occurs instantaneously in many systems (Miranda, Moses, and Heeger 2004; Sheng et al. 2007). This topic has been controversial for a number of years (Yan et al. 1995). Very recently, experiments have been extended to block copolymers (Sreearunothai et al. 2006). Initial theoretical work on excited state absorptions and photoluminescence from interchain species has been reported by Wang, Mazumdar, and Shukla (2007).

Photophysics of metallic SWCNTs (M-SWCNTs). In contrast to typical metals, a nonzero energy gap exists between the second highest valence band and the second lowest conduction band (the $E_{22}$ excitation), and optical absorption across this energy gap is possible. Recent theoretical work has claimed that the lowest excitation across this gap is excitonic (Deslippe et al. 2007). The calculated exciton binding energy (for M-SWCNTs with diameters ~1 nm) is, however, ~0.05 eV, which is considerably smaller than the binding energy of Ex2 (~0.4 eV) in S-SWCNTs with comparable diameters. Experimental work by Wang, Cho, et al. (2007) on the (21,21) M-SWCNTs claims to support this theoretical estimate. Very recent theoretical work by Wang, Fuentes, and Mazumdar, based on the $\pi$-electron model, finds, however, much larger binding energy in the M-SWCNTs (smaller by about 10–15% than S-SWCNTs with comparable diameters). Clearly, more research is warranted here. Experimental ultrafast pump–probe spectroscopy, in particular, will be very useful.

## Acknowledgments

We acknowledge fruitful collaborations and discussions with Professors Z. V. Vardeny and G. Lanzani, and with C.-X. Sheng. Much of the work on PCPs reported here was done with M. Chandross. The work on the $k^1A_g$ states in PPP and PPV was done with A. Shukla and H. Ghosh. Work at the University of Arizona was supported by NSF-DMR-0705163. Work at the University of Hong Kong was supported by RGC grant (HKU 706707P) and Seed Funding of HKU.